\documentclass[preprint2,times]{aastex6}
\pdfoutput=1 
\usepackage{amsmath,amstext}
\usepackage[T1]{fontenc}
\usepackage{apjfonts} 
\usepackage[figure,figure*]{hypcap}
\usepackage{subfigure}
\usepackage{enumitem}
\usepackage{hyperref}
\usepackage{verbatim}
\usepackage{mathrsfs}
\usepackage{afterpage}
\usepackage{color,soul}
\usepackage{marginnote}
\usepackage{longtable}

\newcommand\kh[1]{{[\color{purple} \it KH: #1]}} 
\newcommand*{\al}{\color{red}} 

\newcommand\bb[1]{{[\color{blue} \bf BB: #1]}}  

\renewcommand{\sout}[1]{\ignorespaces}
\renewcommand{\bb}[1]{\ignorespaces}
\renewcommand{\kh}[1]{\ignorespaces}
\renewcommand{\al}[1]{\ignorespaces}

\shorttitle{Statistical tracing of magnetic fields}
\shortauthors{Yuen et al.}

\begin{document}
\title{Statistical tracing of magnetic fields: comparing and improving the techniques}
\author{Ka Ho Yuen\altaffilmark{1}, Junda Chen\altaffilmark{1,2}, Yue Hu\altaffilmark{1,3}, Ka Wai Ho\altaffilmark{4},
A. Lazarian\altaffilmark{1}, Victor Lazarian\altaffilmark{1} Bo Yang\altaffilmark{1}\\ Blakesley Burkhart\altaffilmark{5},
Caio Correia\altaffilmark{7},Jungyeon Cho\altaffilmark{6}, Bruno Canto\altaffilmark{7}, J. R. De Medeiros\altaffilmark{7}}
\altaffiltext{1}{Department of Astronomy, University of Wisconsin-Madison, Madison, WI, USA}
\altaffiltext{2}{Department of Computer Science, University of Wisconsin-Madison, Madison, WI, USA} 
\altaffiltext{3}{College of Electronics and Information Engineering, Tongji University, Shanghai, China}
\altaffiltext{4}{Department of Physics, The Chinese University of Hong Kong, Hong Kong}
\altaffiltext{5}{Smithonian Center for Astrophysics (CfA), Harvard University, Cambridge, MA, USA}
\altaffiltext{6}{Department of Astronomy and Space Science, Chungnam National University, Daejeon, Korea}
\altaffiltext{7}{Departamento de Fisica Teorica e Experimental, Universidade Federal do Rio Grande do Norte, Brazil}

\begin{abstract}
Magnetohydrodynamic(MHD) turbulence displays {\toreferee velocity anisotropies} which reflect the direction of the magnetic field. This anisotropy has led to the development of a number of statistical techniques for studying magnetic fields in the interstellar medium. In this paper, we review and compare three techniques that use radio position-position-velocity data for determining magnetic field strength and morphology : the correlation function anisotropy (CFA), Principal Component Analysis of Anisotropies (PCAA), and the more recent Velocity Gradient Technique (VGT). We compare these three techniques and suggest improvements to the CFA and PCAA techniques to increase their accuracy and versatility. In particular, we suggest and successfully implement a much faster way of calculating non-periodic correlation functions for the CFA. We discuss possible improvements to the current implementation of the PCAA. We show the advantages of the VGT in terms of magnetic field tracing and stress the complementary nature with the other two techniques.
\end{abstract}
\section{Introduction}

Turbulence is {\toreferee an} ubiquitous {\toreferee phenomenon} in astrophysics (see \citealt{D2009})  {\toreferee and it has been detected in the ISM ranges} from kilo-parsecs to sub-AU scales \citep{1995ApJ...443..209A,2004ARA&A..42..211E} {\toreferee and }responsible for the non-thermal broadening of line emission \citep{2013A&A...549A..53K,2014ApJ...785L...1C}.
{\toreferee I}t is a well established fact that the interstellar medium (ISM) is turbulent and magnetized (\citealt{P2004,burkhart10,F2011,VS11}, see \citealt{MK04,MO07} for reviews). Magnetic turbulence controls {\toreferee a number of} key astrophysical processes, e.g. cosmic ray propagation (see \citealt{Sch10}), heat transfer \citep{NM01, L06} {\toreferee and} transfer of polarized radio emission \citep{D05,Hetal06}. {\toreferee Moreover,} ISM {\toreferee t}urbulence and magnetic fields are key components of the star formation paradigm \citep{MO07,P2004,Bialy2017ApJ...843...92B,BB18}. {\toreferee Scientists have} known for decades, for example, that magnetic fields can control the collapse of molecular clouds \citep{MS56,Sp78,Shu83,M91} and remove angular momentum from accretion disks (see \citealt{K12}). More recently, magnetic turbulence has been identified as a driver of magnetic field diffusion from collapsing clouds and accretion disks via the process termed "reconnection diffusion" (see \citealt{LEC17,Metal17}). 

{\toreferee T}he importance of magnetic fields and turbulence {\toreferee has} resulted in {\toreferee the} development of a number of techniques for studying these phenomena {\toreferee in} observations. In general, there exist observational techniques (e.g. polarization or Zeeman studies) in the radio to optical wavelengths to study interstellar magnetic fields (see \citealt{C12} for a review). Techniques based on the statistical imprints of turbulence/gravity have also been suggested and employed (\citealt{Hetal08,L09,Burkhart2017ApJ...834L...1B}, see the thesis \citealt{BB-thesis} for a review). Suggested by the aforementioned statistical works, observational studies should use the feature of turbulence anisotropy when tracing magnetic field directions{\toreferee, e.g. the turbulence anisotropy measured in M51 using the method of correlation function from polarized synchrotron data (Houde et. al 2013).} This opens up {\toreferee a radically new way} to study magnetic fields than the traditional polarization or Zeeman studies.

MHD turbulence {\toreferee has} been explored both theoretically and numerically (see \citealt{Sh83,H84,MM95}. In the present-day, the theoretical foundations of magnetic field tracing {\toreferee techniques} through turbulence statistics are based on the well-known theory proposed by \citeauthor{1995ApJ...438..763G} (GS95, \citeyear{1995ApJ...438..763G}). {\label{somewhere} They developed a theory for strong, incompressible, MHD turbulence that provides definitive predictions of the energy spectrum and anisotropy of velocity fields. GS95 proposes that there is a critical balance between nonlinear interactions and wave propagation, such that the timescales to transfer energy along the two directions are comparable. For an energy-conserving cascade, GS95 implies:
\begin{align}
L_{\parallel} \propto L^{ \frac{2}{3} }_{\perp}
\end{align}}
The above relation is not available in the global system of reference. Therefore one should not expect the anisotropic relation can be observed in the global system of reference. The theory of turbulent reconnection \citep{LV99} {\toreferee has} demonstrated the deep relation{\toreferee ship} and inter-dependence between MHD turbulence and magnetic field dynamics. The framework of \cite{LV99} allows one to understand why, unlike the original GS95 treatment, the anisotropy of turbulence reflects not the mean magnetic field direction, but the direction of magnetic field that percolates turbulent eddies. Indeed, the turbulent reconnection theory predicts that magnetic field lines reconnect so fast that eddies are not constrained by magnetic field if they perpendicularly mix in the direction of the magnetic field of the eddy. This result was confirmed by numerical simulations \citep{CV00,MG01,2003MNRAS.345..325C} and suggested that the study of anisotropy not only can define the mean magnetic field directions, but can also trace local variations of magnetic field direction.


The first suggestion to study magnetic fields statistically using the theoretical understanding of GS95 was the correlation function analysis (CFA) of the velocities \citep{LPE02,EL11,2014ApJ...790..130B}. In the aforementioned papers, the CFA analysis was applied to velocity channel maps obtained from MHD simulations\footnote{Fluctuations of the intensity in so-called "thin velocity channel maps" are mostly influenced by velocity fluctuations, which the meaning of "thin channel" is quantified in \citealt{LP00}).}. It was also demonstrated that the velocity anisotropies can indeed provide the direction of the mean magnetic field. In \cite{EL05} and \cite{EL07} the CFA was quantified and elaborated. The technique {\toreferee was} further explored as a way not only to find magnetic field direction, but also to determine magnetization \citep{EL11,ELP15} as well as to determine the contribution of the fast, slow and Alfven modes in observed turbulence \citep{KLP16,KLP17a,KLP17b}.

The Principal Component Analysis of Anisotropies (PCAA)\footnote{We use PCAA to distinguish this analysis from the earlier studies in \cite{BH02} where they used the Principal Component Analysis (PCA) to get the spectral {\toreferee indices} of turbulence.} provides another way of tracing magnetic field using the turbulence anisotropy  \citep{Hetal08}. The PCAA was successfully applied to the observations and shown to correspond to the polarimetry data \citep{Hetal08}, as well the directions that were obtained with the technique.

Finally, the latest statistical technique for magnetic field studies, the Velocity Gradient Technique(VGT), was demonstrated as a tool to trace magnetic field{\toreferee s} in interstellar medium and molecular clouds. The first work \citep{GCL17}  on VGT used the velocity centroid gradients (VCGs) to trace magnetic field. Only approximate tracing was available and the accuracy of the technique was resolution dependent. A radical improvement of the VGT was achieved in \cite{YL17a}, where the procedure of block averaging was used to provide reliable magnetic field tracing. The further development of the VGT for centroids was done in \cite{YL17b} suggesting {\toreferee that} removing some wavemodes can improve the accuracy of magnetic field tracing. Another branch of the VGT, namely, {\toreferee Velocity Channel Gradients (}VChGs{\toreferee ) } was developed in \cite{LY18a}. {\toreferee The} gradients of intensity fluctuations in thin channel maps were used to represent velocity fluctuations (see \citealt{LP00}). In the same paper \citep{LY18a} {\toreferee suggested} to use the {\toreferee galactic rotation curve} in order to obtain the 3D distribution of magnetic fields. With the numerical studies of velocity gradients in self-absorbing media \citep{GLB17} and application of the VCGs and VChGs to {\toreferee observed neutral Hydrogen (HI)} and {\toreferee molecular tracer maps} \citep{YL17a,YL17b,LY18a} the VGT was identified as a powerful new approach to tracing magnetic fields.


While all these three techniques appeal to GS95 as {\toreferee their} foundation, it is not yet clear whether their prediction{\toreferee s} of magnetic field directions are in agreement with each other. Common questions of comparing these three field-tracing techniques would be:  {\it (1) What are the constraints of the techniques? (2) How precise can we trace the B-field?  (3) Are the methods self-consistent?}. In short, a benchmark study of all three methods in the same framework has yet to be preformed. \cite{YL17a} first showed that VGT is superior over the CFA technique {\toreferee in} tracing magnetic fields in observational data. This result has also been verified on a parallel work using the gradients of Synchrotron Intensities \citep{Letal17}. They point out that, compared to VGT, CFA requires a larger area to perform the ensemble average in calculating the correlation function.  The empirical nature of PCAA also brings questions to its applicability, i.e., there is no self-consistent check for whether PCAA is working in a certain region. VGT shows that having a number of $20^2$ samples is sufficient to satisfy the Gaussian condition showed in \cite{YL17a,YL17b}.  

This paper aims {\toreferee to compare} the three techniques and {\toreferee quantify} their ability to trace magnetic fields (equivalently, detecting anisotropy) using {\toreferee synthetic maps generated from} numerical simulations. {\toreferee As PCAA is not applicable to studying anisotropy in individual channels, we do not show the results of the VChGs analysis, although this technique provides the best tracing of magnetic fields among the different versions of the VGT.}  We investigate the advantages, limitations and constraints of these method{\toreferee s}. We organize the paper as follows: In \S \ref{sec:sim} we describe the details and properties of the simulations. In \S \ref{sec:methodintro} we introduce the three methods of tracing magnetic field in detail. In \S \ref{sec:comp} we show the result of comparison and in \S \ref{sec:discussion} we present a discussion of the results. We conclude our paper in \S\ref{sec:conclusion}

\section{Synthetic Data from MHD Turbulence Simulations}
\label{sec:sim}
The numerical data that we analyzed in this work is obtained by 3D MHD simulations using a single fluid, operator-split, staggered grid MHD Eulerian code ZEUS-MP/3D (Hayes et al. 2006) to set up a three-dimensional, uniform, and isothermal turbulent medium. Periodic boundary conditions are applied to emulate a part of the interstellar cloud. Solenoidial turbulence injections are employed. Our simulations employ various Alfv\'enic Mach numbers $M_A=V_L/V_A$ and sonic Mach numbers $M_S=V_L/V_S$, where $V_L$ represents the injection velocity, $V_A$  the Alfven velocities, $V_S$ the sonic velocity. All the cubes related to this work are listed in Table \ref{tab:sim}. The ranges of $M_S$, $M_A$ and {\toreferee$\beta= 2M_A^2/M_S^2$} are specifically selected so that they cover different possible scenarios of astrophysical turbulence from subsonic to supersonic cases. However, limited by the turbulence scaling (See LV99),  we devote most of our research to the sub-Alfvenic and trans-Alfvenic cases in this study.

To reduce the complexity of comparing the three methods, we only consider the optically thin case and synthesize observational maps using the following treatment. We first compute the PPV cubes from 3D numerical simulations. A PPV cube corresponds to a three-dimensional array with size $n_x, n_y, n_v$, where $n_x, n_y$ represents the sizes along $x$ and $y$ axes, and $n_v$ the number of velocity channels along the spectral line direction v (line of sight, LOS). The number of velocity channels is an adjustable parameter, and in our simulation, we choose $n_v=400$ for our simulation and observation. The {\bf Velocity Centroid} map $C(x,y)$ is a map weighted by velocity channel speed and has the size of $n_x\times n_y$. It is obtained by multiplying each velocity channel by its velocity, and then integrating along the velocity direction, and dividing by the total emission on the direction of integration: 
\begin{align}
C(x,y)=I^{-1}(x,y)\int dv\ v\rho(x,y,v)
\end{align}
where $I$ represents the {\it integrated} intensity of the spectroscopic cube{\toreferee , and $\rho$ is the density of PPV cube}:
\begin{align}
I(x,y)=\int dv\ \rho(x,y,v)
\end{align}

In our implementations for the three methods below, most of our calculations are based on either the velocity centroid $C(x,y)$\footnote{\toreferee We cannot use the latest version of VGT (e.g. Lazarian \& Yuen 2018a, Hu et. al 2018) to compare with either CFA or PCAA. Since in \cite{LY18a} they are performing per-channel gradient studies. For PCAA, it has no ability to perform per-channel prediction of magnetic field. While for CFA only the linearly summed channel anisotropy (Esquivel et. al 2015) is studied instead of the per-channel study for gradients in Lazarian \& Yuen 2018a. To have a fair comparison, we have to use the Yuen \& Lazarian (2017a) version of VGT, which uses the full PPV cube information to compare with CFA and VGT. } or the $\rho(x,y,v)$ (henceforth $\rho$ when the meaning is clear in the context).

\begin{table}
\centering
\begin{tabular}{c c c c c}
Model & $M_S$ & $M_A$ & $\beta=2M_A^2/M_S^2$ & Resolution \\ \hline \hline
Ms0.4Ma0.04 & 0.41 & 0.04 & 0.02 & $480^3$ \\
Ms0.8Ma0.08 & 0.92 & 0.09 & 0.02 & $480^3$ \\
Ms1.6Ma0.16 & 1.95 & 0.18 & 0.02 & $480^3$ \\
Ms3.2Ma0.32 & 3.88 & 0.35 & 0.02 & $480^3$ \\
Ms6.4Ma0.64 & 7.14 & 0.66 & 0.02 & $480^3$ \\ \hline
Ms0.4Ma0.132 & 0.47 & 0.15 & 0.2178 & $480^3$ \\
Ms0.8Ma0.264 & 0.98 & 0.32 & 0.2178 & $480^3$ \\
Ms1.6Ma0.528 & 1.92 & 0.59 & 0.2178 & $480^3$ \\ \hline
Ms0.4Ma0.4 & 0.48 & 0.48 & 2 & $480^3$ \\
Ms0.8Ma0.8 & 0.93 & 0.94 & 2 & $480^3$ \\ \hline
Ms0.132Ma0.4 & 0.16 & 0.49 & 18.3654 & $480^3$ \\
Ms0.264Ma0.8 & 0.34 & 1.11 & 18.3654 & $480^3$ \\ \hline
Ms0.04Ma0.4 & 0.05 & 0.52 & 200 & $480^3$ \\
Ms0.08Ma0.8 & 0.10 & 1.08 & 200 & $480^3$ \\ \hline \hline
\end{tabular}
\caption{\label{tab:sim} Description of the MHD simulation cubes.  $M_s$ and $M_A$ are the instantaneous values at each the snapshots are taken. }
\end{table}

The orientation of anisotropy/gradients from the three methods are compared with synthetic polarization assuming a constant emissivity dust grain alignment process. In other words, the Stokes parameters {\toreferee $Q(x,y)$ and $U(x,y)$} can be expressed in terms of the angle $\theta$ between the y and z direction magnetic fields by $\tan\theta(x,y,z)=B_y(x,y,z)/B_z(x,y,z)$:
{\toreferee 
\begin{align}
Q(x,y)\propto\int dz\rho(x,y,z)\cos(2\theta(x,y,z))\\
U(x,y)\propto\int dz\rho(x,y,z)\sin(2\theta(x,y,z))
\end{align}
}
The dust polarized intensity $I_P=\sqrt{Q^2+U^2}$ and angle $\Phi=0.5atan2(U/Q)$ are then defined correspondingly. 
The alignment between the prediction of magnetic field (CFA,PCAA,VGT) and projected magnetic field {\toreferee orientations} (polarization angles) are {\toreferee quantified} by the \textbf{Alignment Measure (AM)}, introduced in analogy with the grain alignment studies: 
{\toreferee 
\begin{align}
AM=\langle2\cos^2\phi-1\rangle
\end{align}
}
(see \citealt{GCL17,YL17a}), where the {\toreferee $\phi$} is the relative angle between the vectors representing respectively the magnetic field orientation predicted by the three methods and polarization. AM ranges between $-1$ and $1$. When $AM \sim 1$, it means that the two vectors are statistically perfectly aligned; and when $AM \sim -1$, it means that the two vectors are essentially perpendicular to each other. 

\section{Improving the CFA and PCAA}
\label{sec:methodintro}
In this section, we propose improvements for the CFA and PCAA techniques. In particular, for the CFA technique, we propose and test a fast method for calculating the correlation function in non-periodic regions. We also demonstrate {\toreferee a new} way of contour-tracing to significantly improve the accuracy of the CFA technique. 

We also modify PCAA in order to trace magnetic field{\toreferee s} with higher accuracy. To do this, we borrow the block averaging approach that was successfully employed earlier in \cite{YL17a} with velocity gradients. We implement the improvements we suggested for CFA to correlation functions of centroids. Our approaches of improving the calculation of correlation functions are also applicable to the analysis of the velocity channel maps. 

\subsection{Correlation functions from Velocity Centroids}
\label{subsec:CFA}
\begin{figure}
\centering
\label{fig:aniso1}
\includegraphics[width=0.49\textwidth]{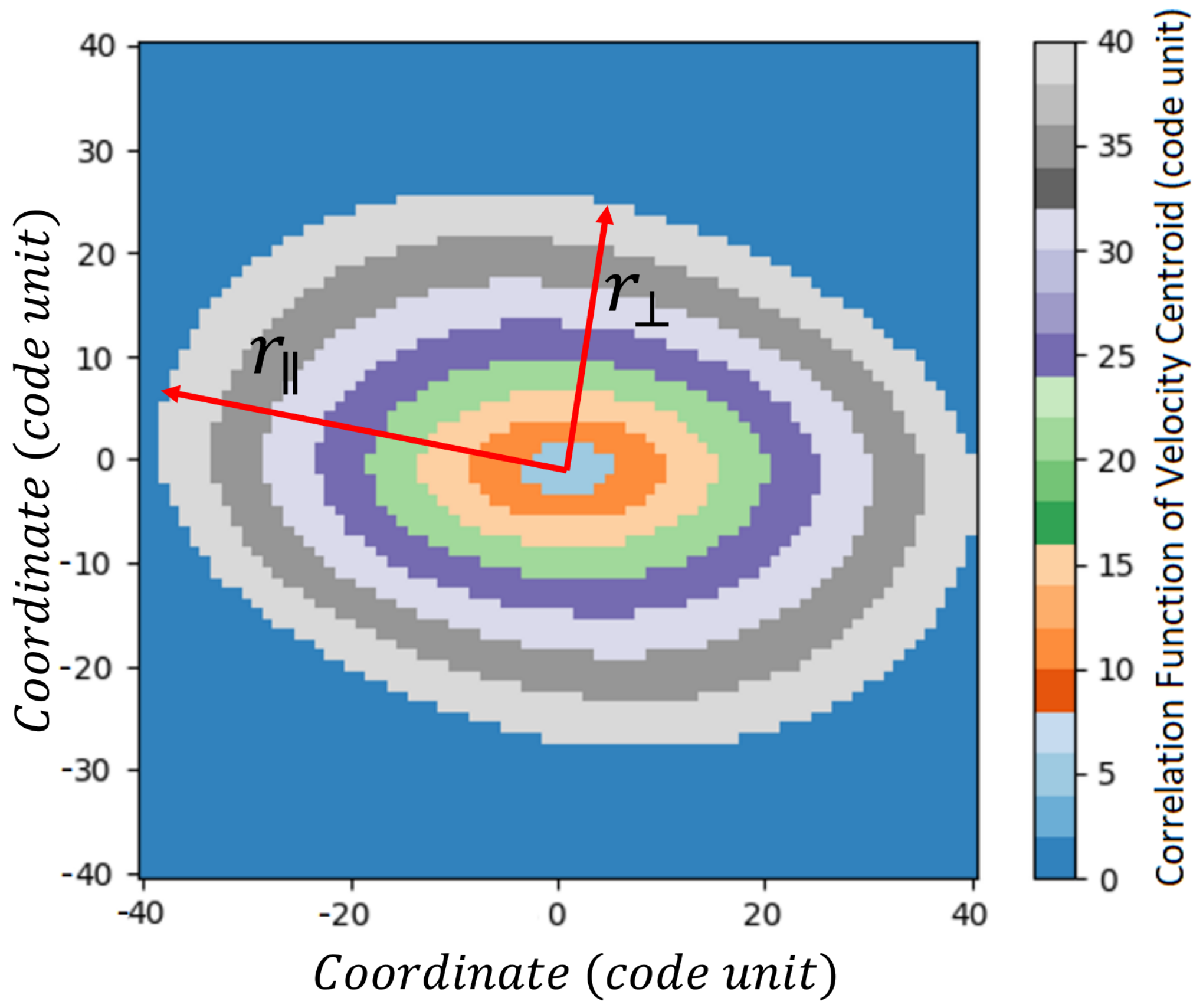}
\caption{An illustration showing how {\toreferee to obtain the correlation function from velocity centroid }(i.e. Eq \ref{eq:fftc}). {\toreferee The correlation function here is calculated from} cube Ms3.2Ma0.32.}
\end{figure}

The (second order) correlation function of a velocity centroid map is defined as:
\begin{align}
\label{eq:cf}
CF_C(\textbf{R})=\langle C(\textbf{r})C(\textbf{r}+\textbf{R})\rangle
\end{align}

The direction of the major axis determines the orientation of {\toreferee the} {\it averaged} magnetic field in a sampled region \citep{EL11, 2014ApJ...790..130B}. For the case of periodic boundary condition simulations, a special form of correlation function and the periodicity of centroid map allow us to obtain the correlation function through the cross-correlation theorem and the fast Fourier transform (FFT): 
\begin{align}
\label{eq:fftc}
CF_C(\textbf{R})= \mathfrak{F}^{-1} \{|\mathfrak{F}\{C\}|^2\}
\end{align}
where $\mathfrak{F}$ is the Fourier transform operator. Figure \ref{fig:aniso1} shows an example of how the correlation function should {\toreferee behave} in terms of the contour plot. One can observe that in {\it scales with the major axis $r_{\parallel}<40$}, the contours are elongated along the mean field direction, and the major axes of the smallest contours are aligned with the parallel direction. Notice that the larger contours are slightly misaligned from the parallel direction  (See \S \ref{sec:discussion} on our explanation of why this is the case).

\begin{figure}
\centering
\label{fig:HEps}
\includegraphics[width=0.49\textwidth]{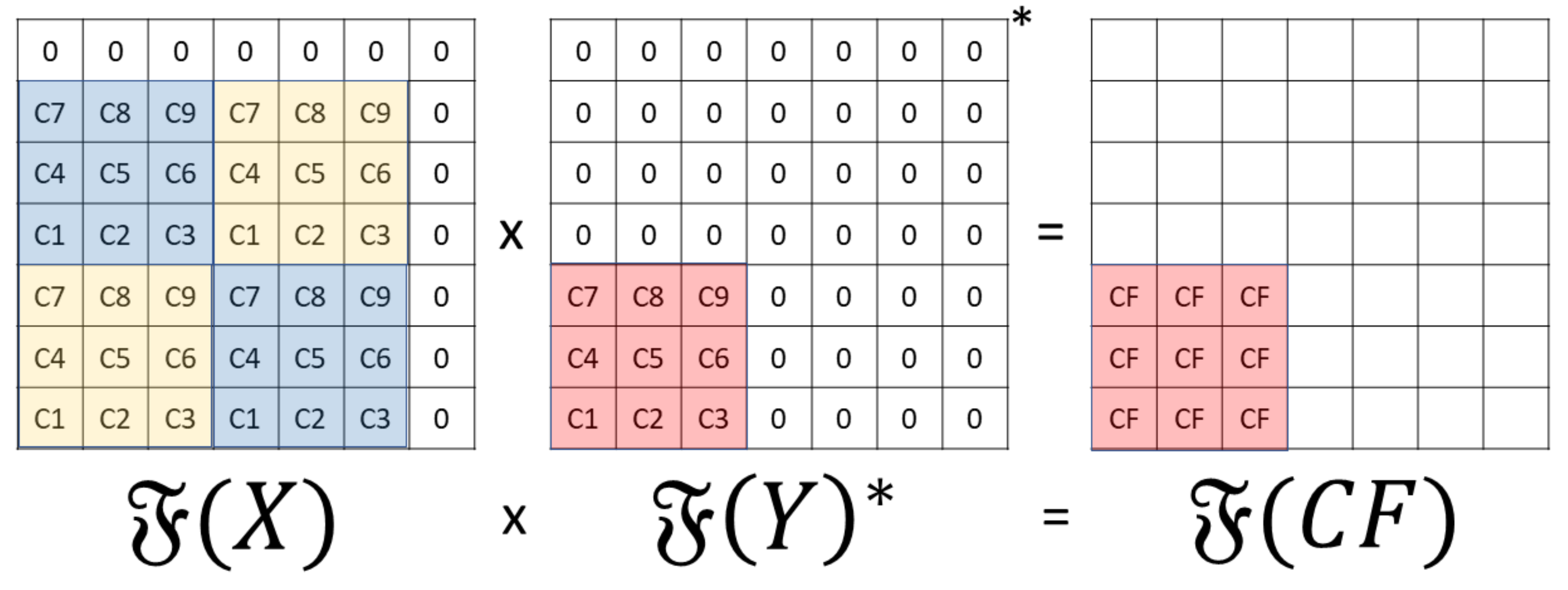}
\caption{An illustration on how {\toreferee to implement the Hockney's approach in our method, i.e. obtain the correlation function from velocity centroid.} (See \S \ref{sec:FFT} for the technical explanation).}
\end{figure}

\begin{figure*}[t]
\centering
\label{fig:CFAillus}
\includegraphics[width=0.98\textwidth]{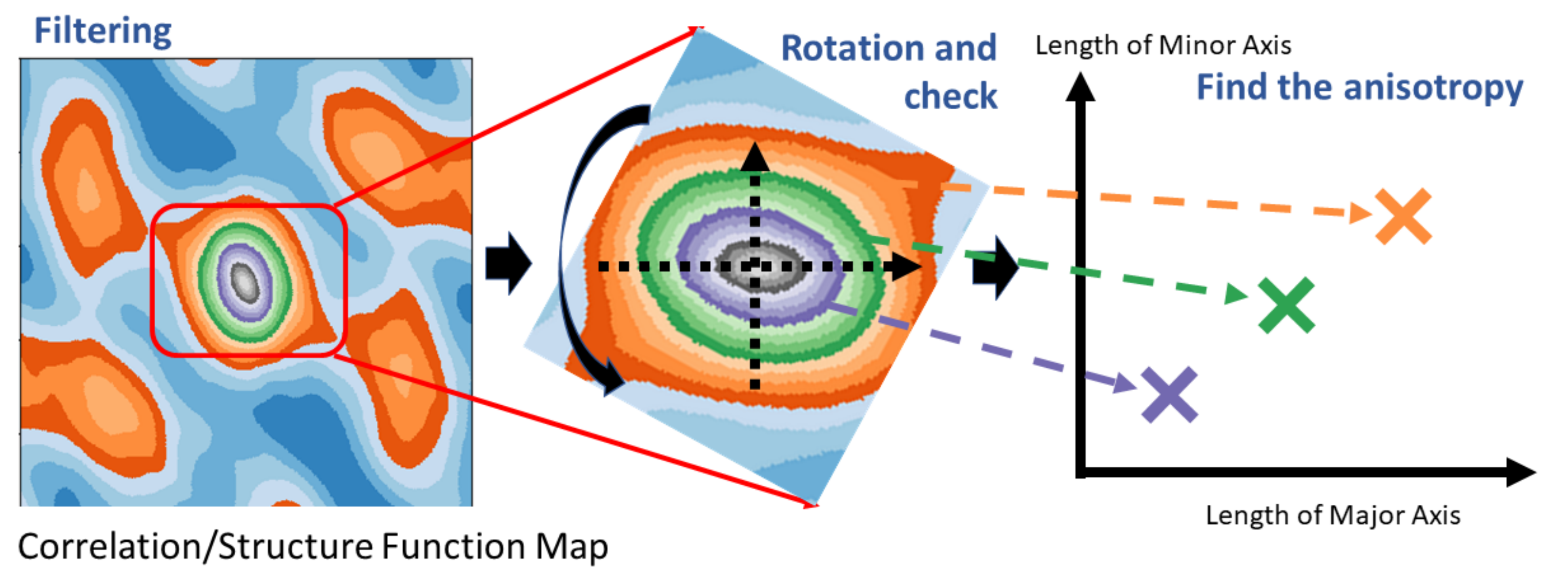}
\caption{An illustration showing how the direction of anisotropy is detected using the rotation-detection algorithm. {\toreferee (Left) We first locate the region having elliptical contours and put the rotation center on the origin of the ellipses. (Middle) Then we slowly rotate the contours so that we identify the major and minor axes and both axes length are recorded for different contours. For example,the big dashed arrows shows the major and minor axes of the dark blue elliptical structure. (Right) The axes are then providing necessarily informations (direction, anisotropic length) for magnetic field studies.}}
\end{figure*}

To compare the CFA with the VGT, we also implement the strategy of block averaging first suggested in \cite{YL17a} to CFA. In the framework of VGT, block averaging reveals the {\it statistically most probable} direction of magnetic field in the region of consideration. Considering the sub-block statistics, computation of correlation function anisotropy should also reveal the direction of magnetic field similar to the block averaging in VGT. 

The implementation of block statistics in CFA requires computing Eq. \ref{eq:fftc} for non-periodic maps. However Eq. \ref{eq:fftc} is limited to maps that are periodic in both boundaries. For non-periodic regions, one needs to compute the correlation function through the direct computation of Eq \ref{eq:cf}, which requires the amount of computation in proportion to the square of the number of pixels. In particular, to utilize CFA inside blocks, which naturally are not periodic, a faster calculation on par with Eq \ref{eq:fftc} should be developed. To calculate the correlation function of an non-periodic centroid map, we adopt the Hockney method (Hockney 1968) to solve the open-boundary convolution problem as shown {\toreferee in} Fig \ref{fig:HEps} formally with Eq \ref{eq:fftc} (See \S \ref{sec:FFT} for a formal discussion)\footnote{Assuming one has N data for centroid, then the traditional method (Eq. \ref{eq:cf}) requires a complexity of $O(N^2)$,with the FFT method (Eq. \ref{eq:fftc}), the complexity reduced to $O(NlogN)$.}, which decrease the time complexity of the computation process. We pad the centroid map C with size $n_x \times n_y$ (in Fig \ref{fig:HEps} $n_x=n_y=3$) into two $(2n_x+1)\times(2n_y+1)$ block $X,Y$ as shown in Fig. \ref{fig:HEps},where 
\begin{align}
X(i,j)&=
\begin{cases}
C(mod(i-1,n_x)+1,mod(j-1,n_y)+1), & 1\leq i,j \leq 2n\\
0, & otherwise
\end{cases}\\
Y(i,j)&=
\begin{cases}
C(i,j), & \qquad 1\leq i,j \leq n_x\\
0, & \qquad otherwise
\end{cases}
\end{align}
where $mod$ is the modulo operation. The open-boundary correlation function is therefore
\begin{align}
\label{eq:obcf}
CF_C(\textbf{R})[i,j]= \mathfrak{F}^{-1} \{\mathfrak{F}\{X\}\mathfrak{F}\{Y\}^*\}, \qquad 1\leq i,j \leq n_x
\end{align}
This implementation enables one to compute the correlation function and structure function efficiently in non-periodic cases, which are particularly useful for our comparison {\toreferee between} CFA {\toreferee and} VGT.\footnote{The Big-O factor for the Hockney's method is $O((2N+1)log(2N+1))$ compared to the traditional method with $O(N^2)$, where $N$ is the number of discrete elements in an array.} Figure \ref{fig:CFAillus} shows how to locate the direction of anisotropy given a specific correlation map or structure function map. Concretely, the algorithm plots the contour lines of the map, and detects the orientation of the elongated major-axes and minor-axes of each (elliptical) contour line. Then the map is rotated such that the major-axis of a contour with particular radius (in our case, the searching radius is $10$ pixels) is parallel to the horizontal direction. The direction of anisotropy is then determined by the direction of major axis of the contour.

\begin{figure}
\centering
\label{fig:rotate-cfa}
\includegraphics[width=0.49\textwidth]{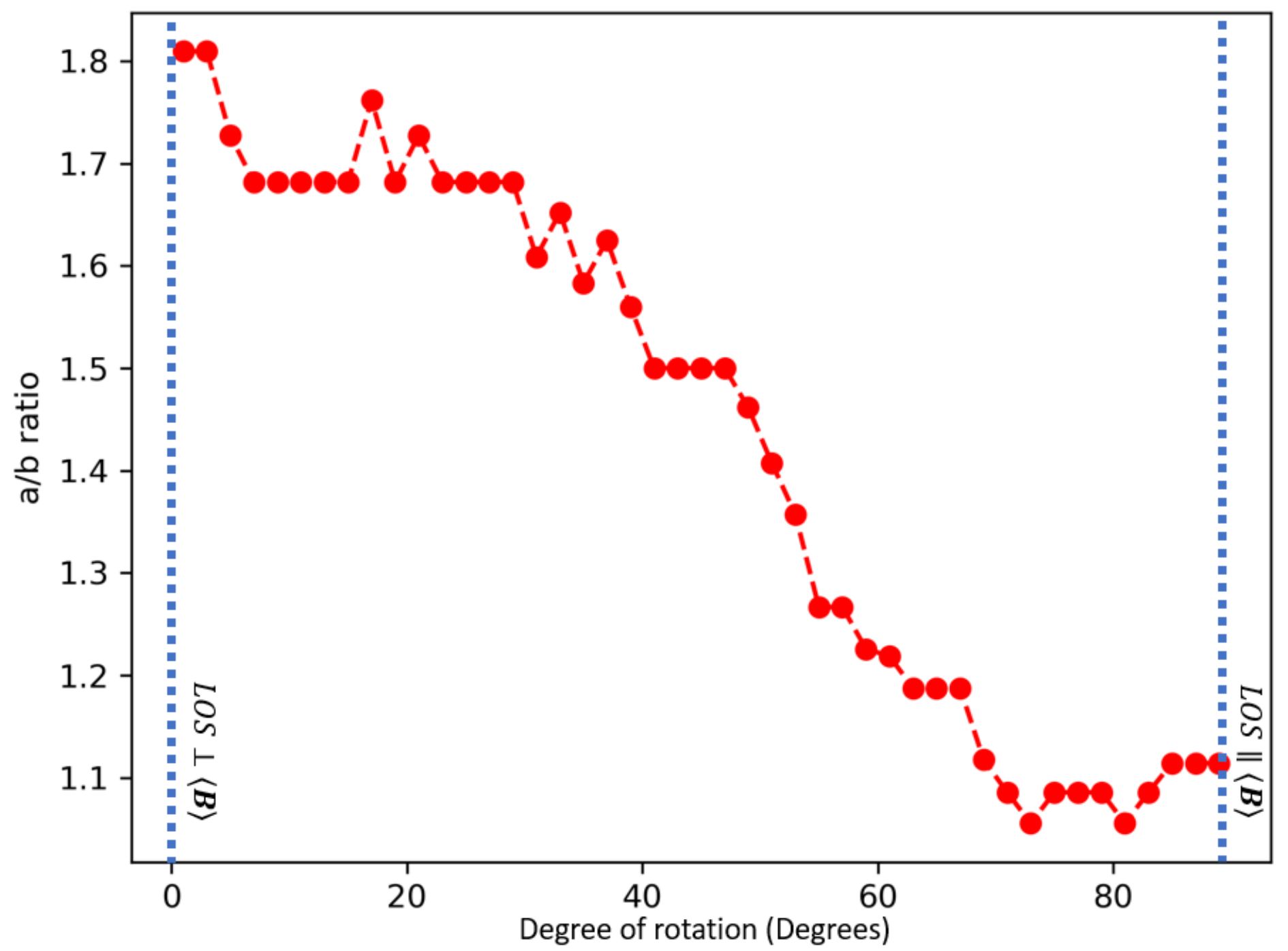}
\caption{{\toreferee The change of a/b ratio with respect to the relative angle between the LOS and the mean magnetic field direction.}}
\end{figure}

There are {\toreferee additional} difficulties in using CFA when the line-of-sight makes a different angle to the mean magnetic field.  \cite{2014ApJ...790..130B} {\toreferee suggested} that the {\toreferee detected} degree of anisotropy will drop when the angle between the line-of-sight to mean magnetic field decreases. We also see the same effect for our numerical cubes as in Fig \ref{fig:rotate-cfa}. For VGT the respective investigation has been done in \cite{YL17b} and having similar drop of alignment measure when the angle between the line-of-sight to mean magnetic field decreases. 
{\toreferee However, observers should be aware of the fact that, while the degree of anisotropy is decreasing with respect to the decrease of angles between line of sight and the mean magnetic field, the predicted orientations are still evident for both VGT \citep{YL17b} and CFA (Fig \ref{fig:rotate-cfa}). }

\subsection{Analysis with PCA}
\label{subsec:PCA}

\subsubsection{Finding anisotropies with PCA}
The PCA is widely used in image processing and image compression. In terms of astrophysical applications, the PCA analysis was used in \cite{2002ApJ...566..276B,BH02}  for obtaining the turbulence spectrum from observations. In \cite{Hetal08} the PCA was employed for studying turbulence anisotropies. The physical meaning of the eigenvalues from the PCA analysis are closely related to the value of the turbulence velocity dispersion $v^2$. In particular, those larger eigenvalues correspond to the largest scale contributions of turbulence eddies along the line of sight $v^2 \sim (l^{1/3})^2 \sim l^{2/3}$, assuming GS95 scaling applies.   

\begin{figure*}
\centering
\label{fig:PCAillus}
\includegraphics[width=0.98\textwidth]{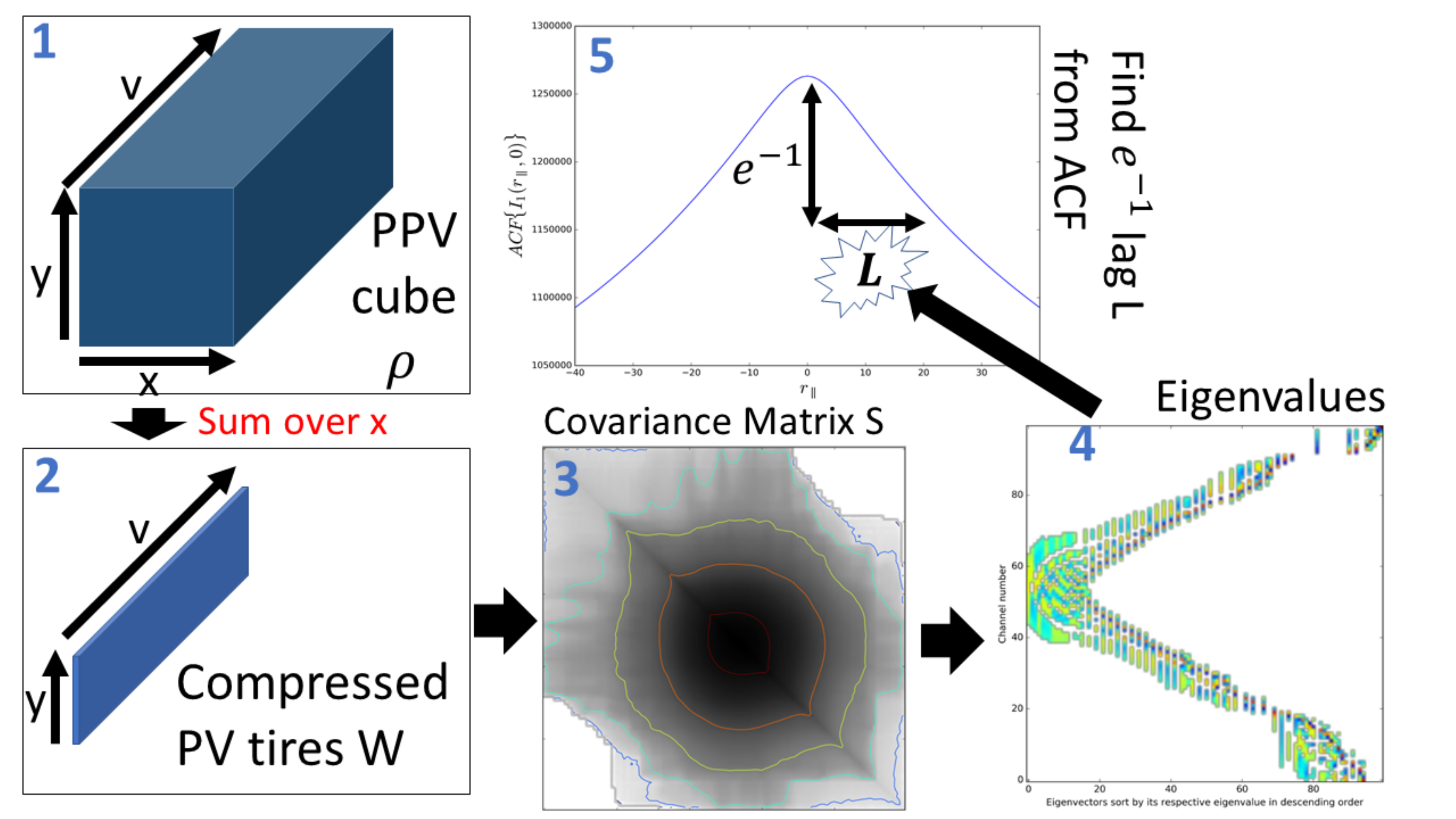}
\caption{{\toreferee An illustration showing the pipelines of PCAA. We first project the PPV cubes along either of the spatial directions (panel 1), which we would call the compressed PPV cubes "PV tires" (panel 2) in which the covariance matrix $S$ of the PV tires (panel 3, Eq. \ref{eq:W}) can be computed. The eigenvalues of the covariance matrix (panel 4) contains the information velocity information along this particular PV tire, and the $e^{-1}$ lag of the autocorrelation functions of the eigen-vectors and eigen-projections (panel 5, Eq. \ref{eq:ACF})  tells the characteristic velocities $\delta v_{x,y}$, and length scales $L_{x,y}$ respectively.}}
\end{figure*}

To study anisotropy, \cite{Hetal08} applied the PCA to the spectroscopic data as a tool of tracing anisotropy, similar to what was done earlier in the statistical analysis of channel maps and centroids \citep{LPE02,EL05}. Similar to the latter techniques, the directional PCAA demonstrated its ability to identify the direction of the mean magnetic field. To help the reader to understand the essence of the technique we provide a simple version of its mathematical formalism as well as a pictorial illustration in Figure \ref{fig:PCAillus}.

Assuming {\toreferee a proper} normalization is used \footnote{In principle one shall use the normalized PPV cube $\rho' = \rho/\int \rho$. However for the treatment of PCAA, the difference of a constant does not alter the result. Therefore we stay with using $\rho$ for simplicity.}, we can treat the PPV cube $\rho(x,y,v)$ as the probability density function of three random variables $x,y,v$. The covariance matrix is:\footnote{The correct definition of covariance matrix should be $S(v_1,v_2)=E(\rho(v_1) \rho(v_2))-E(\rho(v_1))E(\rho(v_2))$, where E is the expectation operator. However the second part was not included in \cite{Hetal08} }
\begin{align}
S(v_1,v_2) \propto \int dxdy \rho(x,y,v_1)\rho(x,y,v_2)
\end{align}

In the later treatment of anisotropy tracing, \cite{Hetal08} splits the Position-Position-Velocity (PPV) cube into vertical and horizontal Position-Velocity tires (PV tires), where every PV tire is a vertical or horizontal slice from the PPV map $\rho(x,y,v)$ averaged over the x-direction (y-direction): 
\begin{align}
W(y,v)\propto\int dx\rho(x,y,v)\\
W(x,v)\propto\int dy\rho(x,y,v)
\end{align}

The covariance matrices ($S_x$ and $S_y$) for the PV tires (W) are
\begin{align}
\label{eq:W}
S_x(v_1,v_2) \propto \int dx W(x,v_1)W(x,v_2)\\
S_y(v_1,v_2) \propto \int dy W(y,v_1)W(y,v_2)
\end{align}

hence {\toreferee} eigenvalue equation for {\toreferee these} covariance matri{\toreferee ces are}: 
\begin{align}
S_x\textbf{u}_x=\lambda_x\textbf{u}_x\\
S_y\textbf{u}_y=\lambda_y\textbf{u}_y
\end{align}

where the $\lambda_{\{x,y\},i}$ are the eigenvalues associated with the eigenvectors $\textbf{u}_{\{x,y\},i}$ with $i=1,2,...,n_v$.  The eigenvectors contain the information of velocity variations along this particular PV tire. To get information of the spacial variance, one must project each eigenvector into the PV tires. These eigen-projections, $P_{x,i}, P_{y,i}$ are:  
\begin{align}
\label{eq:P}
P_{x,i}(x)=\int dv W(x,v)\textbf{u}_{x,i}(v)\\
P_{y,i}(y)=\int dv W(y,v)\textbf{u}_{y,i}(v)
\end{align}

With the sets of eigenvectors and eigen-projections on the $x$ and $y$-direction in hands, one can apply the method of autocorrelation functions (ACFs) $ACF\{X\} = CF\{X\}/Var\{X\}$ to these sets of data in order to obtain the characteristic velocity and scale. Each characteristic velocity (scale) is calculated when the ACF for one eigenvector (eigen projection) drops by one e-fold. Due to resolution limitations, the characteristic velocities (scales) are interpolated between the nearest points to $1/e$: 
\begin{align}
\label{eq:ACF}
\frac{ACF\{u\}(\delta v)}{ACF\{u\}(0)}=e^{-1}\\
\frac{ACF\{P\}(L)}{ACF\{P\}(0)}=e^{-1}
\end{align}
We obtain at least 10 pairs (See Right of Fig \ref{fig:PCAreflux}) of characteristic velocity $\delta v_{x,y}$, and scale $L_{x,y}$ from the ACFs of the correspondent eigenvectors and eigen projections.

If, as it is in the case of numerical data cubes, the magnetic field is oriented either {\toreferee along} $x$ or $y$ axes, one can expect the ACFs for $x$ and $y$-directions to be different. When using observational data, \cite{Hetal08} attempted to find the direction of magnetic field by calculating the ACFs while rotating the directions of the $x$ and $y$ axes. This by itself can provide the magnetic field direction. However, \cite{Hetal08} were studying the scaling of the ACFs while changing the orientation of the coordinate axes. The anisotropy was determined by the variations of the exponent $\alpha$ in:
\begin{align}
\label{eq:PCAscaling}
\delta v_x=v_{0,x}L_x^{\alpha_x}\\
\label{eq:PCAscaling2}
\delta v_y=v_{0,y}L_y^{\alpha_y}
\end{align}

An example can be found in Fig \ref{fig:PCA1}. There are some challenges associated with this procedure. Indeed, according to both turbulence theory and  MHD turbulence simulations  (GS95, LV99, \citealt{2003MNRAS.345..325C}) the differences of {\toreferee indices} should not be observed in the global system of reference related to the mean magnetic field. In the Appendix B we show that  the observed differences between the {\toreferee indices are} the result of both the limited inertial range of numerical simulations and the isotropic driving of turbulence at the injection scale.  

While we find the approach in \cite{Hetal08} has problems, for the sake of comparison, we use{\toreferee their formalism on the velocity and length-scale determinations (Eq. \ref{eq:ACF})} as it is presented in their original work. In particular, we find the differences between the exponents $\alpha_x$ and $\alpha_y$. Compared to the actual observational study in which the direction of magnetic field is not known a priori, {\toreferee a rotation of the} coordinate system {\toreferee is required to {\it guess} the direction of magnetic field before applying PCA.}\footnote{The corresponding procedure is not elaborated in detail in Heyer et al. (2008). We feel that this procedure of rotating of the coordinate system is not straightforward in terms of its practical implementation. For instance, the calculation of the covariance matrix $S_x$ for the PV tires (Eq. \ref{eq:W}) requires an addition along the y-axis {\it in the rotated coordinate.} For both synthetic and observation maps, the information of the map is usually stored in a rectangular coordinate. Any kinds of addition after a rotation, as the PCA method did, will result in a distortion in the covariance matrix $S_{x,y}$ }.

\subsubsection{Testing PCAA}

\begin{figure*}[t]
\centering
\label{fig:PCA1}
\includegraphics[width=0.49\textwidth]{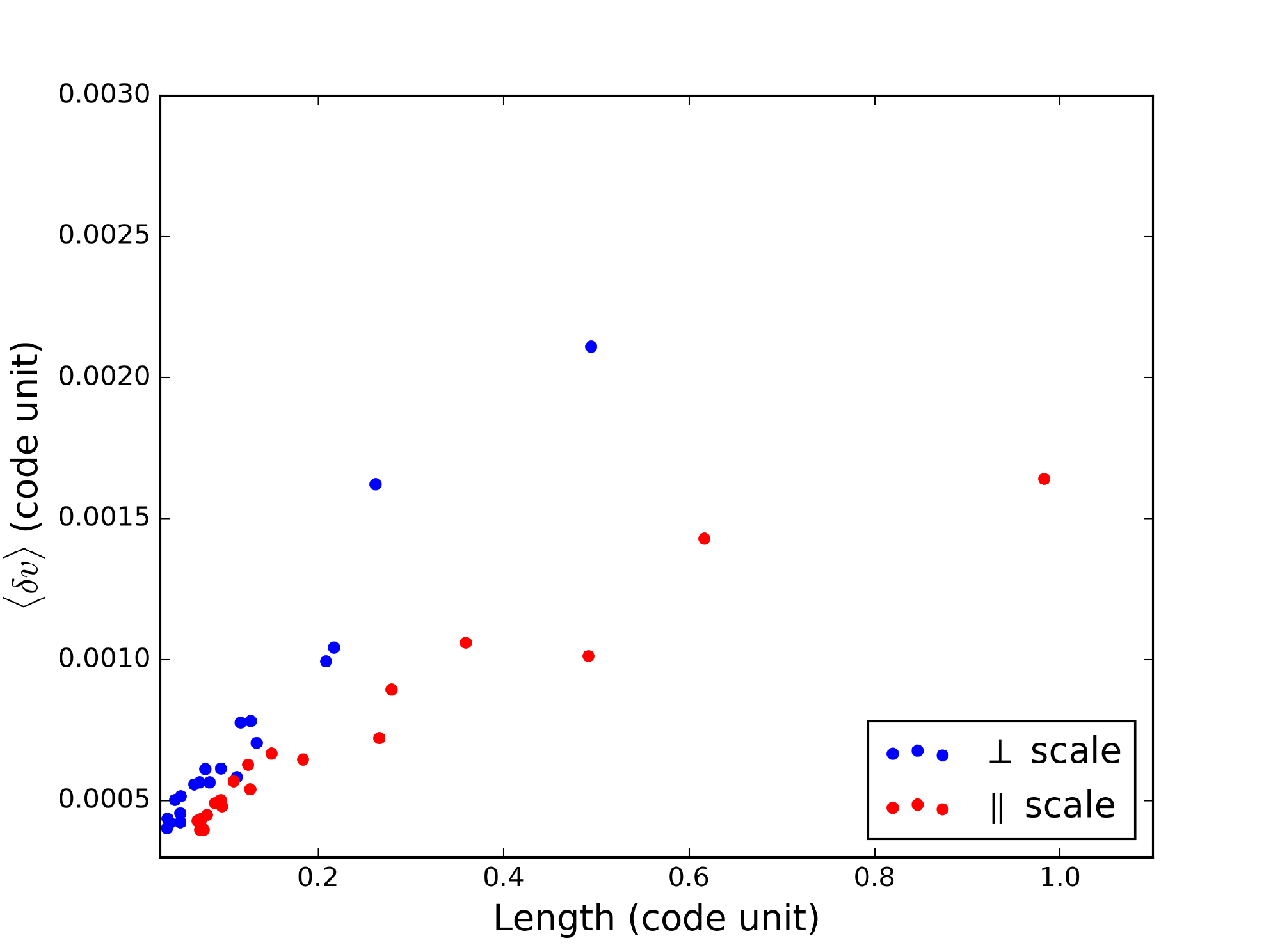}
\includegraphics[width=0.49\textwidth]{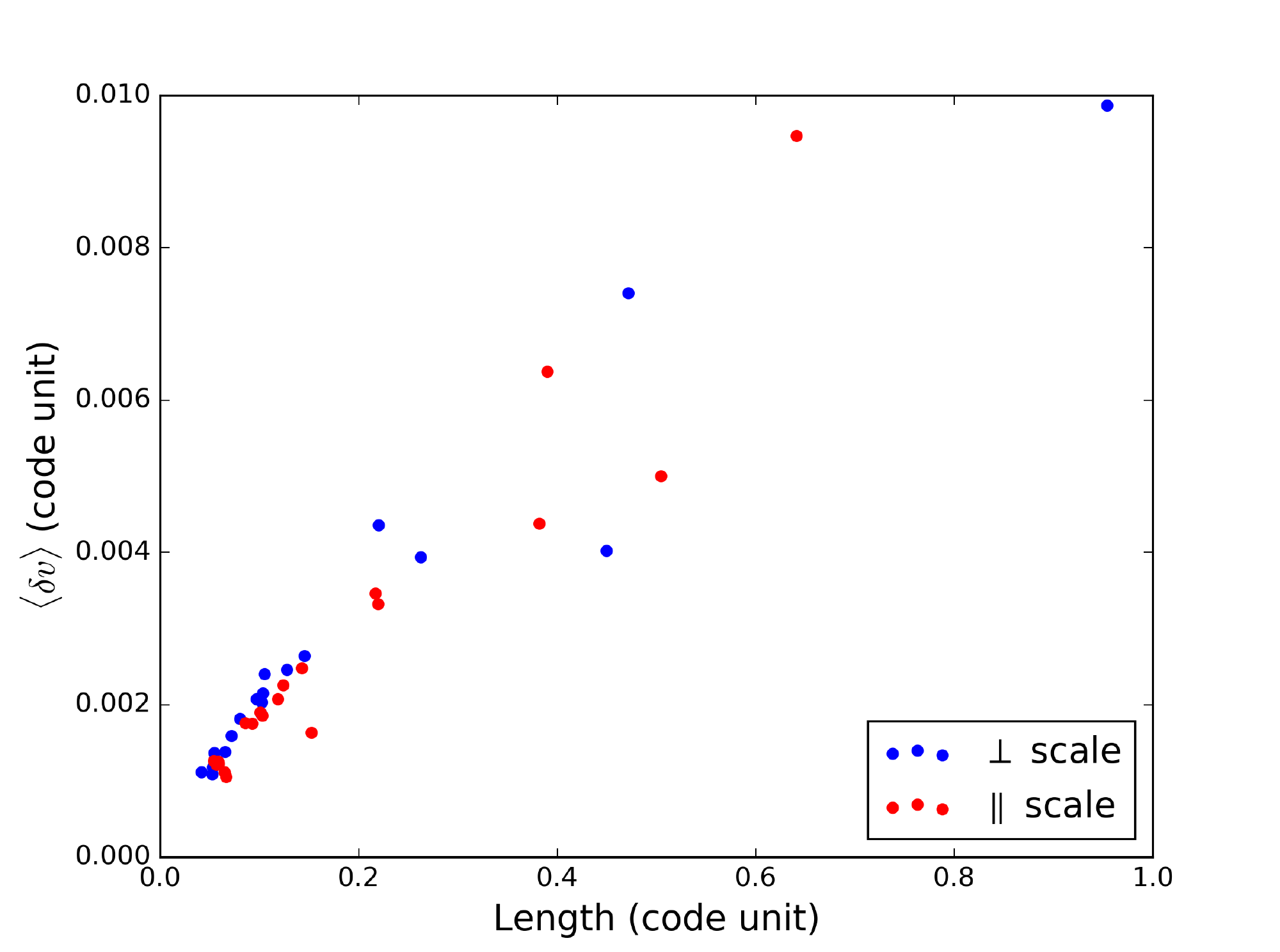}
\caption{\toreferee Two panels of scatter plots showing how the parallel and perpendicular pairs of $(\delta v, L)$ vary for the case when the PPV cube is constant-density (Left) or real-density (Right). }
\end{figure*}

\begin{figure*}[t]
\centering
\label{fig:VGT-PCA1}
\includegraphics[width=0.49\textwidth]{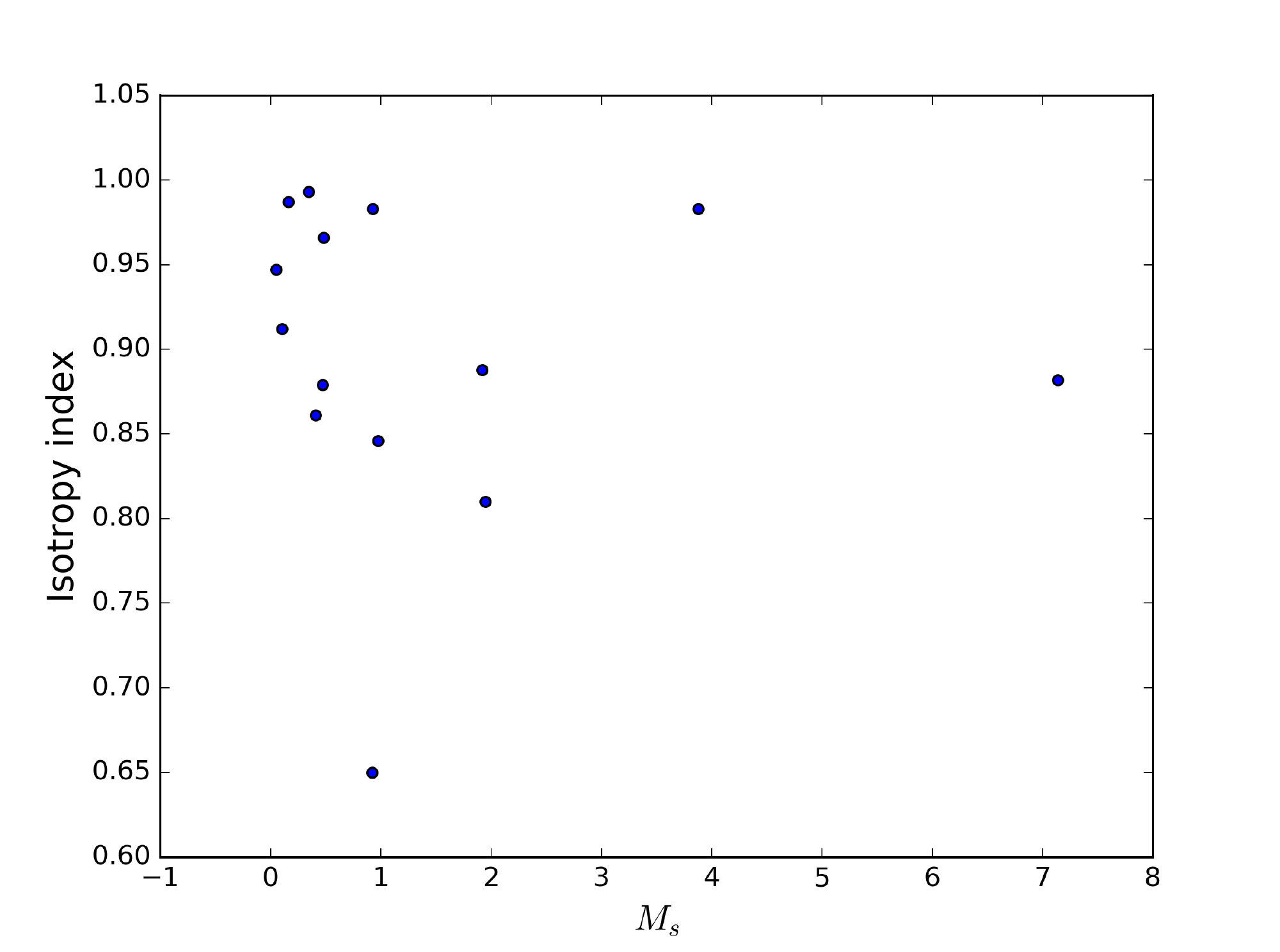}
\includegraphics[width=0.49\textwidth]{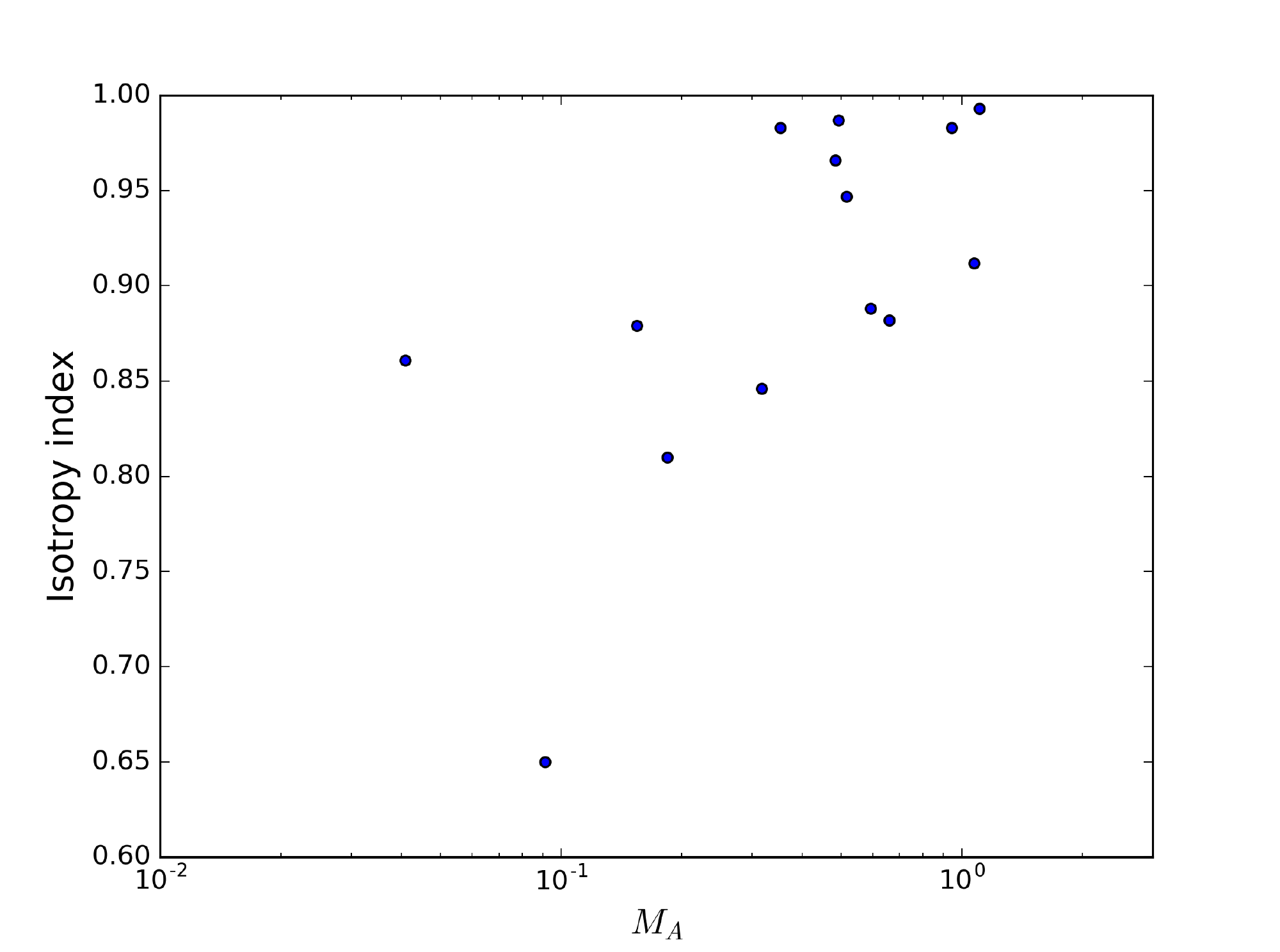}
\caption{{\toreferee The plots show how the isotropy index (y-axis, see Eq (\ref{eq:isoindex})) varies with respect to $M_s$ (Left figure, x-axis) and $M_A$ (Right figure, x-axis) }}
\end{figure*}

We first prepare the PPV cubes with constant $n$ and {\toreferee PPV} density $\rho$ by the distribution function of the line of sight velocity $f(v;z)$
\begin{align}
\label{eq:nppv}
n(x,y,v) &= \int dz f(v;z)\\
\rho(x,y,v) &= \int dz \rho(x,y,z) f(v;z)
\end{align}
We then apply the method of PCAA as illustrated. Fig. \ref{fig:PCA1} shows how the density scaling in PPV cubes would change the anisotropies found by PCAA on the same numerical cube but with different weighting to density.

We also test our implementation of PCAA on our simulations as listed in Table \ref{tab:sim}. Fig \ref{fig:VGT-PCA1} shows how the sonic Mach number $M_s$ and Alfvenic Mach number $M_A$ could possibly change the anisotropy. In this work we  {\toreferee adopt an} \textit{isotropy} index {\toreferee which can be obtained} directly from the PCAA exponents so we can compare to Velocity Centroids isotropy index, 
\begin{align}
\label{eq:isoindex}
\Upsilon=1-\frac{|\alpha_\perp-\alpha_\parallel|}{\sqrt{\alpha_\perp\alpha_\parallel}}
\end{align}
For isotropic velocity fields, $\Upsilon\sim1$. Note that $\Upsilon$ can {\toreferee be} negative for highly anisotropic clouds. 

{\toreferee In} Fig \ref{fig:VGT-PCA1}, we do not see a clear relationship between the isotropy index and $M_s$, but a slightly positive relationship between isotropy index and $M_A$ is found. This is expected as in the PPV formulation using PCAA. only the largest variance contributions are extracted, and it is well known that the variance of density is a function of sonic Mach number $\sigma^2 \propto log(1+b^2M_s^2)$ for some $b\sim 1/3 - 1/2$ \citep{F2011,2012ApJ...755L..19B}. {\toreferee If as mentioned in the previous section only the largest eigenvalue is extracted}, only the density clumps with the highest dispersion will, therefore, be analyzed.

We explained earlier that, in the global system of reference, there should not be a difference of the spectral {\toreferee indices} based on the theory of MHD turbulence. To compare with VGT, which is a local measure of anisotropy, we have to improve the method of PCAA to the local scale instead of a global direction. In \S \ref{subsec:vgtpca} we shall show our method of improving PCAA and compare with VGT.

\subsection{The Velocity Gradient Technique}

\subsubsection{Block averaging}

\begin{figure}[t]
\centering
\includegraphics[width=0.49\textwidth]{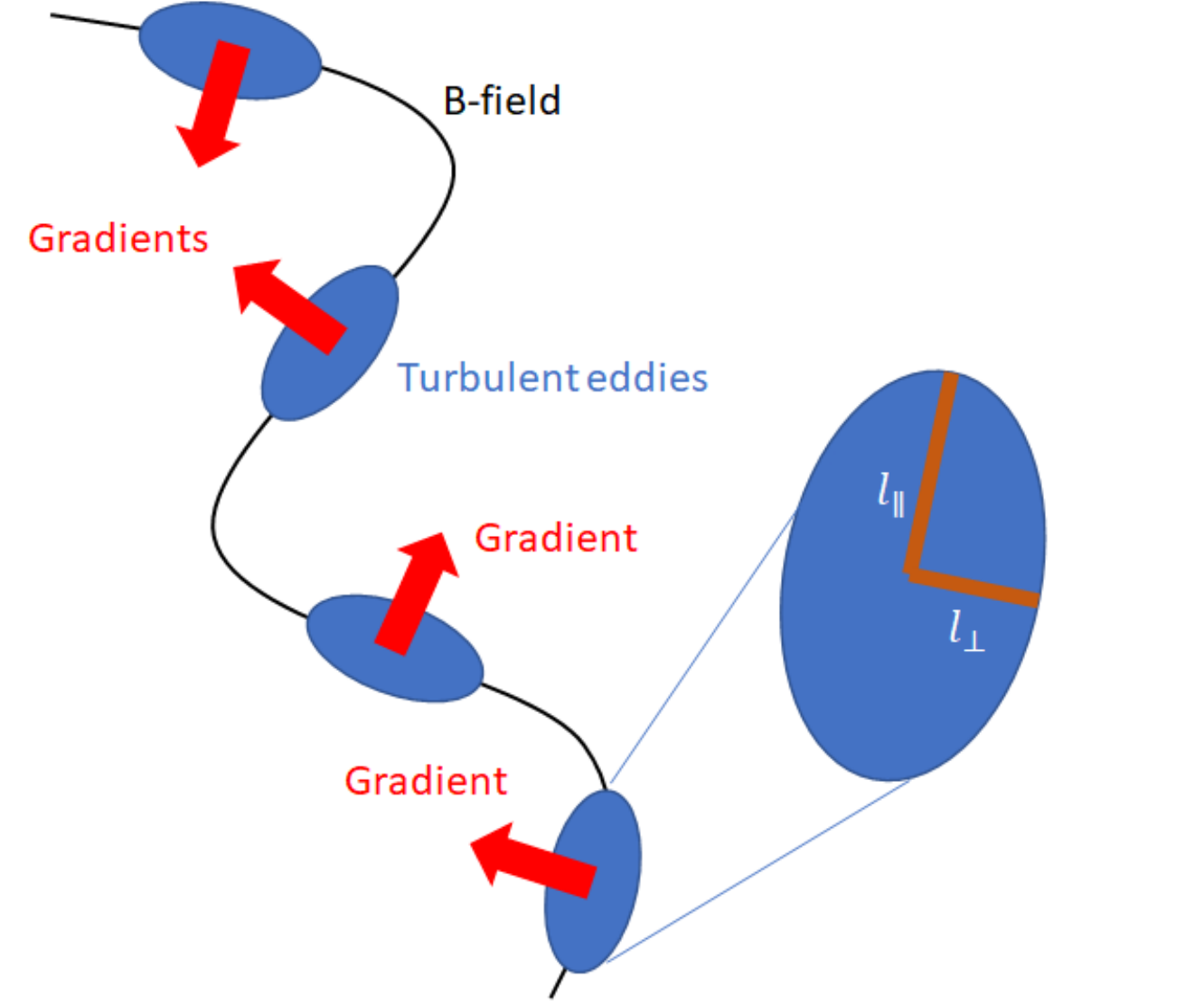}
\caption{\label{fig:vgtillus} An illustration showing how VGT works.{\toreferee The MHD turbulent eddies (blue) are elongated along the local magnetic field (black line) directions. The gradients (red) of physical informations of these eddies (e.g. densities without shock influences, velocities, magnetic intensities) will therefore be maximally perpendicular to the local magnetic field directions. As a result, a simple $90^o$ rotation of gradients trace the direction of magnetic field locally.}}
\end{figure}

The Velocity Gradient Technique is a recently developed technique for tracing magnetic field directions based on the anisotropic turbulence scaling (GL17a, YL17a). In terms of the GS95 scaling, turbulent eddies are elongated along the local magnetic field directions. As a result, the gradients of velocity  are perpendicular to the major axis of anisotropy, and thus the local magnetic field directions {\toreferee (See Fig. \ref{fig:vgtillus} for an illustration)}. In simple words, one can use the gradients of observables (e.g. Intensities I and Centroid C) to estimate the direction of magnetic fields by simply rotating the gradients by $90^o$

We adopt the sub-block {\toreferee averaging} and the {\toreferee respective} error-estimation method as suggested in YL17ab. 
{\toreferee While the gradients are good probes of magnetic field directions as suggested by our series of papers \citep{YL17a,YL17b,LY18a},} knowing the errors of individual gradient vector is always beneficial when applying to observations. We have to emphasize on the basis of GS95 turbulence that the statistical nature of gradients acted similarly to the techniques based on turbulence anisotropy and principle component analysis \citep{EL05,Hetal08,2014ApJ...790..130B}. An insufficient number of pixels within the block will result in a significant error of magnetic field direction estimation. The recipe we proposed in YL17ab allows us to acquire the statistical gradient orientation average within a block from the peak value of the Gaussian fitting function $N(\theta;p_1,p_2,p_3) = p_1\exp(-(\theta-p_2)^2/p_3^2)$ in the gradient orientation distribution. The standard error of the Gaussian peak, $\delta p_2$, which is one of the free parameters of the Gaussian function for fitting, will tell us how good the gradient orientation distribution follows the Gaussian distribution, and how accurate the peak can represent the {\it averaged} direction of gradients inside a particular block of certain size.

\subsubsection{Recent Improvements for the VGT}
{\toreferee Recently, it was demonstrated by \citet{2018arXiv180208772H} that the method of Principal Component Analysis (PCA) is capable in extracting the anisotropic velocity modes along the line of sight. Different from the method of PCAA, \citet{2018arXiv180208772H} construct the eigen-intensity maps $I_{eigen}$ and eigen-centroid maps $C_{eigen}$ using PCA as:
\begin{equation}
C_{eigen}(x,y)=\frac{\int dv\ \rho(x,y,v)\cdot v \cdot \lambda(v)}{I_{eigen}(x,y)}
\end{equation}
\begin{equation}
I_{eigen}(x,y)=\int dv\ \rho(x,y,v)\cdot \lambda(v)
\end{equation}
where the $\lambda$ are the eigenvalues associated with the eigenvectors $\textbf{u}$.

In both synthetic and observational maps, the extraction of eigen-centroids can effectively probe the direction of magnetic field with very high accuracy. As a result, for the studies of the projected magnetic field, the improved technique
can provide higher accuracy of magnetic field tracing.}

\section{Comparison between the three techniques}
\label{sec:comp}

The {\toreferee common} goal of the three methods (CFA, PCAA, VGT) is to trace magnetic field orientation independently from polarimetry measurements. For VGT, the accuracy of determining the magnetic field direction can be obtained through block averaging \citep{YL17a,LY18a}.\footnote{In our forthcoming paper we are comparing the VGT with tracing of filaments in channel maps that is suggested in Clark et al. (2014, 2015). We feel that the filaments in the latter papers are results of the velocity crowding, the same effect that makes thin channel maps sensitive to velocity fluctuations (Lazarian \& Pogosyan 2000).} The other two techniques have their limitations. For instance, the CFA technique depends strongly on the viewing angle chosen \citep{2014ApJ...790..130B}. We have not developed yet a {\toreferee self-consistent} procedure for estimating the accuracy of the magnetic field orientation with CFA {\toreferee like the Gaussian fitting criterion in VGT sub-block averaging \citep{YL17a}}.  The PCAA technique in its present incarnation seems to have even more problems. The determination of the anisotropy angle using PCAA requires tedious checking of anisotropy index in {\it every} possible angle that the map can rotate. Moreover, the projection of PV tires after rotation will result into distortion of the PV tire statistics. As a result, the autocorrelation function may not provide the anisotropy direction correctly. {\toreferee Moreover, the perpendicular and parallel velocity scaling indices $\alpha$ are not expected to change for turbulence with extensive inertial range.}

Nevertheless, the synergy of the techniques should be utilized. We note that an advantage of CFA is that, the related anisotropies are analytically described in \cite{KLP17a} and are related to the contributions from slow, fast and Alfv\'en modes that constitute the MHD turbulence cascade \citep{2003MNRAS.345..325C}. Below we provide a more quantitative comparison of the 3 techniques.

\subsection{VGT vs CFA}
\label{sec:VCTvCFA}
\begin{figure}
\centering
\label{fig:VCG1}
\includegraphics[width=0.49\textwidth]{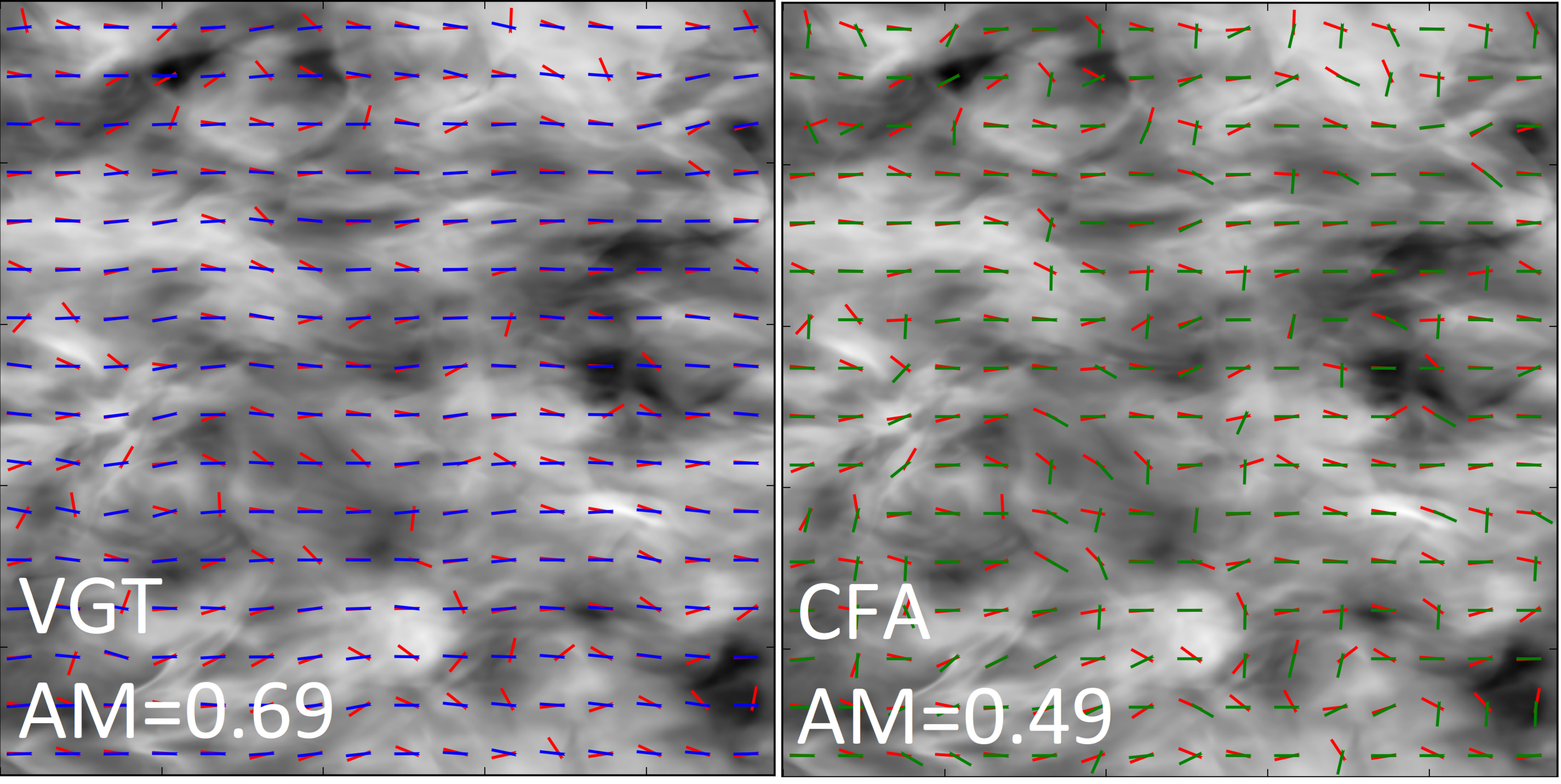}
\caption{A plot showing the B-field orientation probed by VGT (red) and CFA (green) versus the magnetic field directions (blue) overlaid on the velocity centroid maps with both block size of 30 pixels on the cube {\toreferee with  $M_s=1.92$ \& $M_A=0.59$}.}
\end{figure}

Fig. \ref{fig:VCG1} shows a visual comparison of applying both VGT and CFA on the same centroid data from the Ms1.6Ma0.528 simulation by selecting a block size of 30 pixels. We compare the performance of CFA with the older VGT recipe (from \citealt{YL17a}), which the latter carries only one user-defined variable, the block size. For the B-field orientation probed by CFA, we use the orientation of the major axis as a prediction following the treatment in \S \ref{subsec:CFA} (See Fig. \ref{fig:CFAillus}). The algorithm suggested in \S \ref{subsec:CFA} is more adaptive in dealing with irregular anisotropic shapes compared to our previous treatment in \citep{YL17a} using a highly simplified gradient-at-origin for correlation function method\footnote{In principle, the shorter axis direction for the anisotropy corresponds to larger gradients. Therefore one can try to detect the anisotropy by taking gradients at the origin of correlation functions and rotate $90^o$ for magnetic field directions. This, however, cannot tackle complex structures like what we see in Fig \ref{fig:CFAdis3}}. We select the pixel distance of 10 pixels for anisotropy contour detection.

\begin{figure}
\centering
\label{fig:VCG2}
\includegraphics[width=0.49\textwidth]{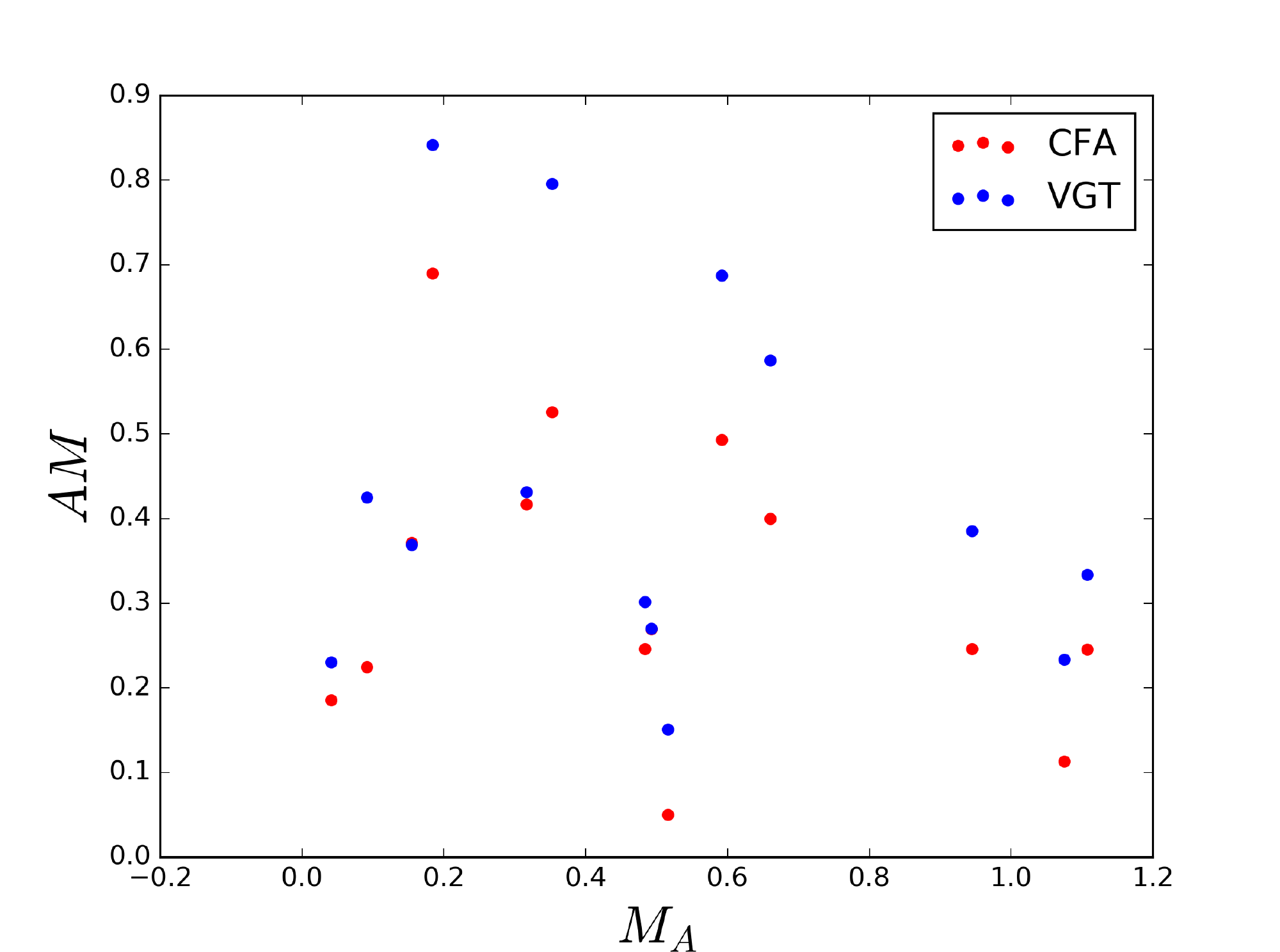}
\caption{A scatter plot showing the AM of VGT (Blue) and CFA (Red) to B-field respectively against the Alfvenic Mach number $M_A$. }
\end{figure}

One can see a significant advantage of VGT compared to CFA in Fig. \ref{fig:VCG1} in terms of the alignment measure. Figure \ref{fig:VCG2} shows a scatter plot of $AM_{VGT}$ to $AM_{CFA}$ with respect to $M_A$ using the gradient recipe of \cite{YL17a} for a block size of 30 pixels. The mean AM for VGT is $\sim 0.43$ while that for CFA is about $\sim 0.31$. While we do not see a clear trend of {\toreferee $\Delta AM=AM_{VGT}-AM_{CFA}$} versus $M_A$, there is a general of $\Delta AM=0.2$ advantage for VGT over CFA. 

We do expect that the tracing power of VGT increase appreciably after the improvements suggested from \cite{LY18a} .But whether we can use the same improvement technique for CFA-probed B-field is not clear yet. One can visually see {\it from Fig \ref{fig:VCG1}} 
that the magnetic field estimations from CFA are more likely to be bi-model, i.e. the vectors are likely to be either parallel or perpendicular to the real field. This might due to the fact that small-scale statistical shape studies are not well studied (See Fig \ref{fig:CFAdis3} for a pictorial illustration).

\subsection{VGT vs PCAA}
\label{subsec:vgtpca}
To compare PCAA with VGT, we update the implementations of PCAA to make it comparable to the sub-block averaged method in VGT. The general guideline would be {\it performing PCAA on sub-blocked PPV cubes}. We extract the partial PPV (pPPV) cubes $\rho_{ij}$ that covers partial spatial regions:
\begin{align}
\rho_{ij}(x,y,v) = \rho((i-1)n+x,(j-1)n+y,v)
\end{align}
The sequence of PPV cubes $\rho_{ij}$ contains in total $n_x n_y/n^2$ elements. For each pPPV cube, we assume they are independent and processed using the steps from \S \ref{subsec:PCA}. One can refer to Figure \ref{fig:PCAillus} for the simplified, pictorial work flow for PCAA. The product from the pipeline would be a 6-element array $(\delta v_x,\delta v_y,L_x,L_y,\alpha_x,\alpha_y)_{ij}$ for each pPPV cube, which they should have an empirical scaling as shown in Eq (\ref{eq:PCAscaling}) \& (\ref{eq:PCAscaling2}) .

One can try to convert the anisotropic direction predicted by PCAA in each PPV cube \footnote{This is, however, not a very accurate statement, as PCAA has a pre-assumed anisotropy direction. In \cite{Hetal08} they illustrate how the anisotropy direction by rotating the Taurus map and see which orientation can give the largest anisotropy difference. } to some magnetic field orientation prediction similar to VGT. The trick is to use the fact that the 6-element array provides a measure of velocity gradients, and the maximal-perpendicular properties of velocity gradients allow us to predict the direction of magnetic field by rotating PCA-backed velocity gradient by $90^o$. We start with the {\it statistical average} maximal PCAA-gradient orientation inside the block $(i,j)$ assuming both $L_x,L_y$ are small and not aligned with the $\parallel,\perp$ coordinate
\begin{align}
\nabla v(L_x,L_y) \sim \langle \frac{\delta v_x}{L_x},\frac{\delta v_y}{L_y} \rangle
\end{align}
The orientation of PCAA gradient is then given by,
\begin{align}
\label{eq:PCAangle}
\tan\theta(L_x,L_y) = \frac{\delta v_yL_x}{\delta v_xL_y} =\frac{L_y^{\alpha_y-1}}{L_x^{\alpha_x-1}}
\end{align}

For $L$ smaller than some turbulence scales depending on the Alfvenic Mach number (See LV99 for a complete discussion){\toreferee, i.e. $L$ is sufficiently small,} we {\toreferee can then} obtain an expression from Eq \ref{eq:PCAangle} of approximating the PCAA angle:
\begin{align}
\label{eq:pcaaangle}
\tan\theta = \frac{1-\alpha_x}{1-\alpha_y}
\end{align}

\begin{figure*}[t]
\centering
\label{fig:VGT-PCA-C}
\includegraphics[width=0.49\textwidth]{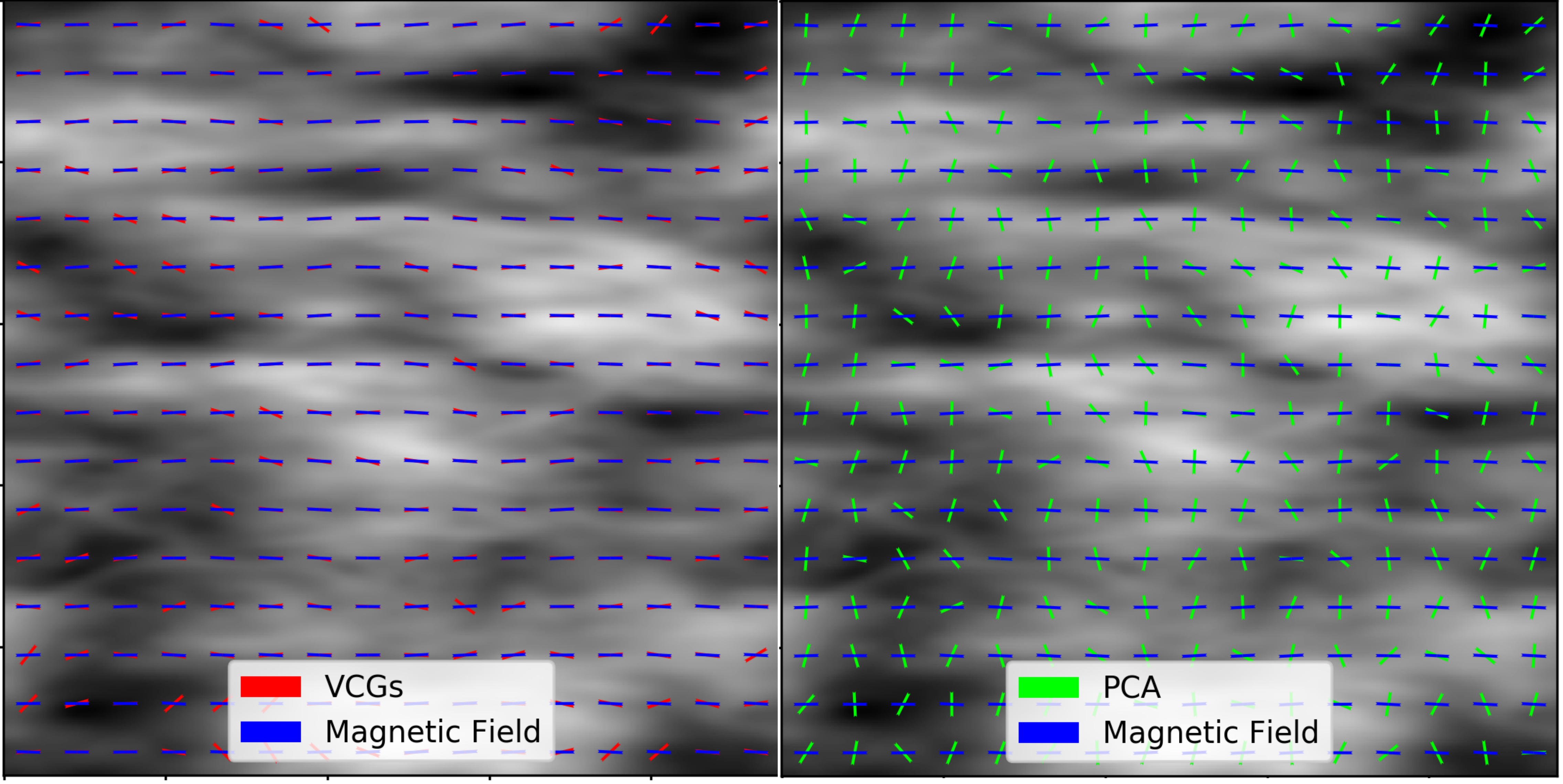}
\includegraphics[width=0.49\textwidth]{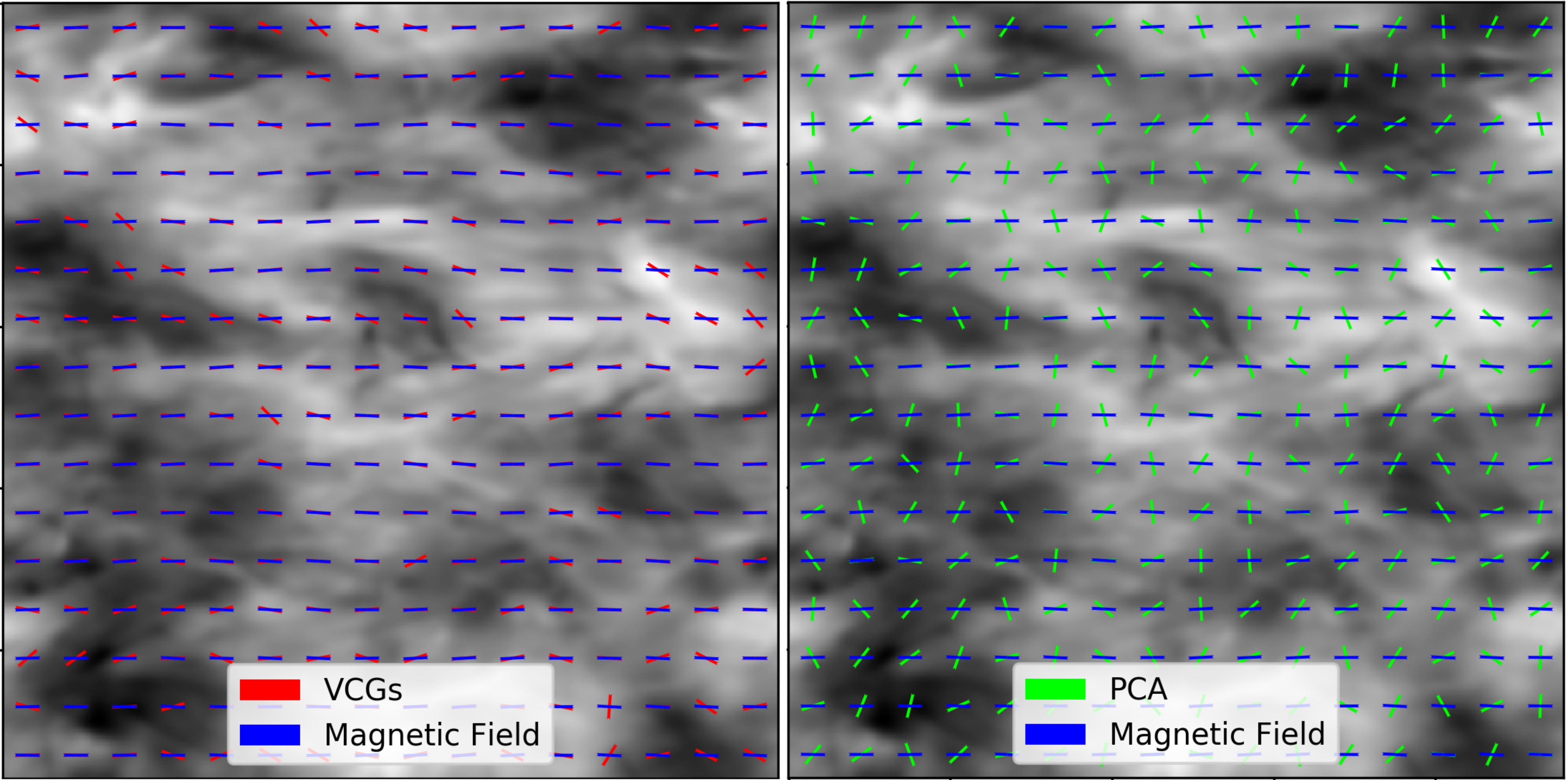}
\caption{A comparison between the magnetic field predictions from VGT and PCAA when the constant density condition (first and second panel) and turbulent real density condition (third and fourth panels) are applied respectively. In the first two panels, the fluctuations are entirely due to velocity fluctuations, since the density is held constant. In this figure we use the Model Ms3.2Ma0.32. Block size$=30$.}
\end{figure*}

Fig. \ref{fig:VGT-PCA-C} shows how the magnetic field orientation predicted by  $\theta+\pi/2$ from Eq. (\ref{eq:pcaaangle}) is compared to polarization measurements and VGT. In the following, we discuss the two separate cases regarding properties of PPV cubes.

\subsubsection{Constant Density Case}

We first investigate how the block-averaging would behave in the constant density PPV cube \citep{2013ApJ...770..141B}. In other words, we create PPV cubes with a uniform density field and a turbulent velocity field.  Thus all fluctuations in such a cube are entirely due to velocity caustics. On the left of Figure \ref{fig:VGT-PCA-C} shows how the performance of PCAA is compared to VCG visually in a centroid map from super-Alfvenic simulation Ms3.2Ma0.32. For some part of the region, PCAA is able to trace magnetic field accurately; however, the alignment measure for the PCAA-magnetic field is negative, which means the results from PCAA are almost perpendicular to magnetic field. Comparatively, VCG has an alignment measure almost close to one. This suggests that VCG is more accurate for tracing the direction of the magnetic field. 

\subsubsection{Real Density Case}

We also explore how the use of real density PPV cube would change the alignment measure of VGT and PCAA. In theory the involvement of real density will make the velocity channel map contains both density and velocity contribution, in which the proportion among the two contribution is determined by the channel thickness (LP00) and also the sonic Mach Number. When we sum up the channels, density is expected to dominant over velocity contribution, and the alignment measure is expected to drop compare to the constant density case. 

On the right of Fig \ref{fig:VGT-PCA-C} shows how the performance of PCA is compared to VCG visually in a centroid map from super-Alfvenic simulation Ms3.2Ma0.32. Comparing with the constant density case, the alignment of PCA is improved closing to zero. However, VCG still provides an excellent performance in tracing magnetic field, even through we do see a drop of alignment measure from the constant density case. 

\section{Additional Effects}
\label{sec:discussion}
\subsection{Fluctuation of anisotropy scale and directions in CFA}

The method of CFA has its limitations in both resolution and quality of the data, both to simulations and observations. In principle, numerical simulations have limited inertial ranges. In terms of correlation and structure functions, only the small-scale contributions are considered to be meaningful. This is also true for observation where bulk motions exist (e.g. galactic motions and shear, outflow) aside from turbulence. As a result, the large-scale part of the correlation function may not be so meaningful in determining the anisotropy. However, this immediately brings a paradox of the anisotropy direction, as the small-scale part is often limited with only several pixels only and highly depends on the quality of the data.\footnote{\toreferee While readers might challenge whether the large-scale shearing/rotation motion or the small-scale outflow motions may alter the result of VGT. In principle, with proper scale filtering \cite{YL17b} one can remove the contribution from large scale structures, which is also true for the method of CFA. For the small scale outflow motion using a large enough block size can average out the contribution of non-turbulent motions.} {\toreferee If the resolution of the map is small} (in observation) or the dissipation process is strong (in numerical studies), the small-scale anisotropy determination under the assumption of elliptical elongation would fail (See Sec 3.1 for the method building). The change of anisotropy is even more severe when the number of samples for the statistical studies is not enough (e.g. Sec 4.1). To what extent the correlation function anisotropy can provide a correct answer given a map with certain resolution is uncertain.

\begin{table}[h]
\centering
\begin{tabular}{c c c c c}
Model & $M_S$ & $M_A$ & $\beta=2M_A^2/M_S^2$ & Resolution \\ \hline \hline
H0 & 7.36 & 0.22 & 0.0017 & $792^3$ \\
H1 & 6.41 & 0.41 & 0.0083 & $792^3$ \\
H2 & 6.47 & 0.61 & 0.0176 & $792^3$ \\
H3 & 6.47 & 0.80 & 0.0309 & $792^3$ \\
H4 & 6.15 & 1.00 & 0.0531 & $792^3$ \\ \hline
\end{tabular}
\caption{\label{tab:highres} Description of the MHD simulation cubes with larger resolution.  $M_s$ and $M_A$ are the instantaneous values at each the snapshots are taken.  }
\end{table}

We therefore want to test the dependences of resolution to the anisotropy method using multiple resolutions. We prepared some higher resolution cubes (Table \ref{tab:highres}) and compare with what we have (Table \ref{tab:sim}) for both the anisotropy axis ratio and its orientation.

\begin{figure}
\centering
\label{fig:CFAdis0}
\includegraphics[width=0.49\textwidth]{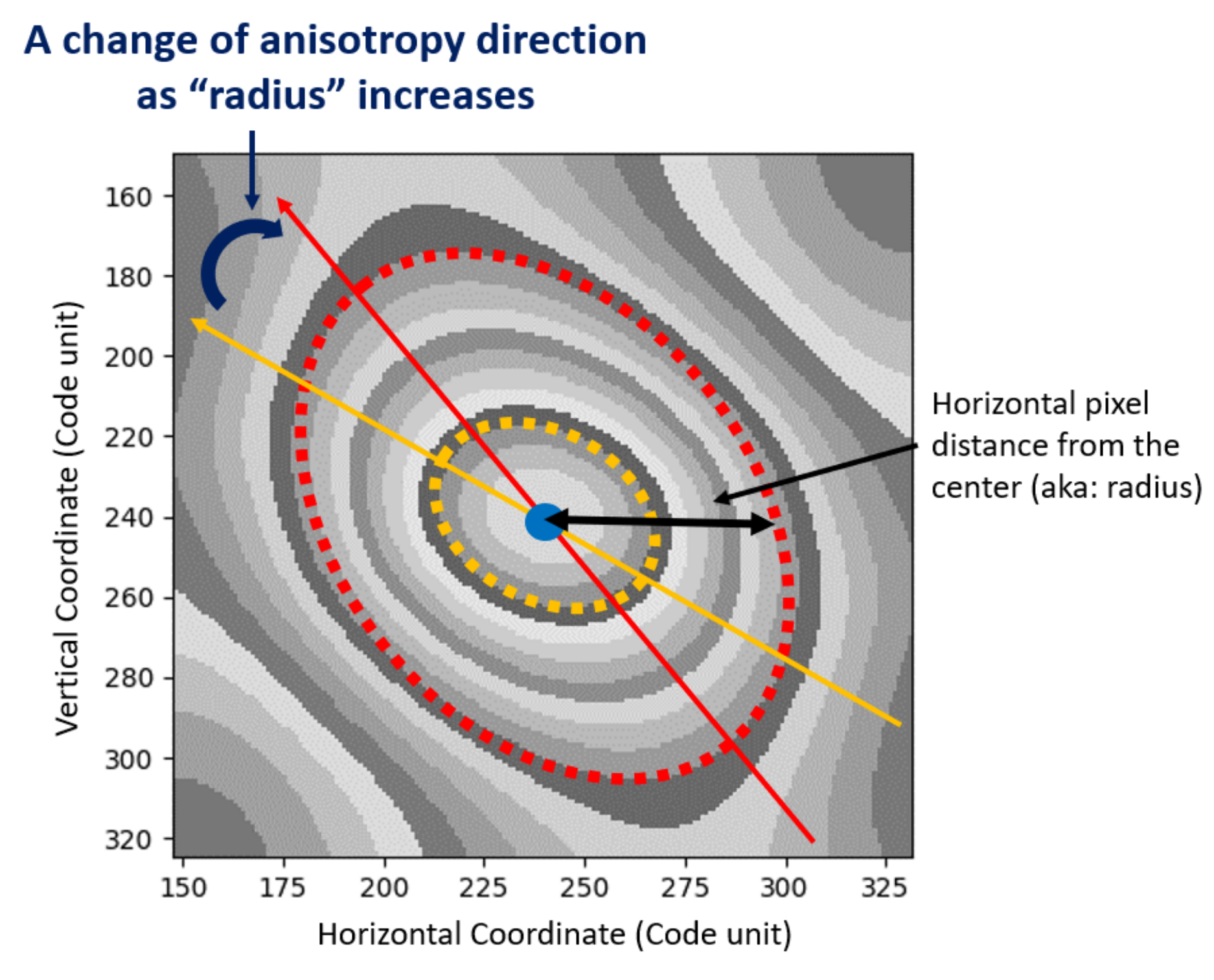}
\caption{An illustrative figure showing how the direction of anisotropy changes with respect to contour size: A yellowish contour with "radius" (Defined in the figure as the horizontal pixel distance from the center) of $\sim 20$ is selected and the direction of anisotropy for this contour is shown by a yellow arrow. One can pick a bigger contour (e.g. the reddish one we picked with "radius" of $\sim 60$), the direction of anisotropy for this contour is changed significantly. }
\end{figure}

\subsubsection{Distortion of anisotropy over scales}
\begin{figure*}[t]
\centering
\label{fig:CFAdis1}
\subfigure[]
{\includegraphics[width=0.49\textwidth]{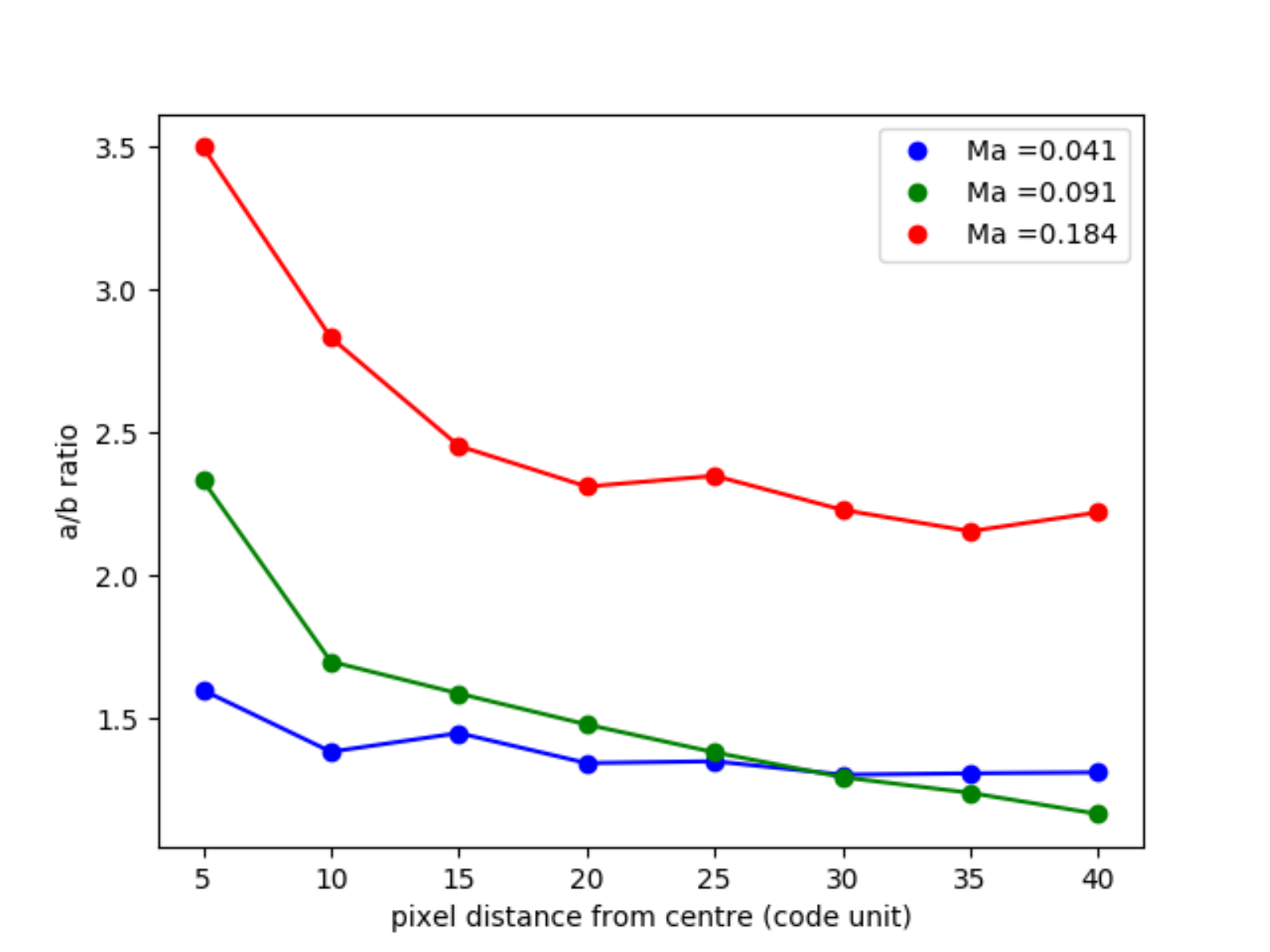}}
\subfigure[]
{\includegraphics[width=0.49\textwidth]{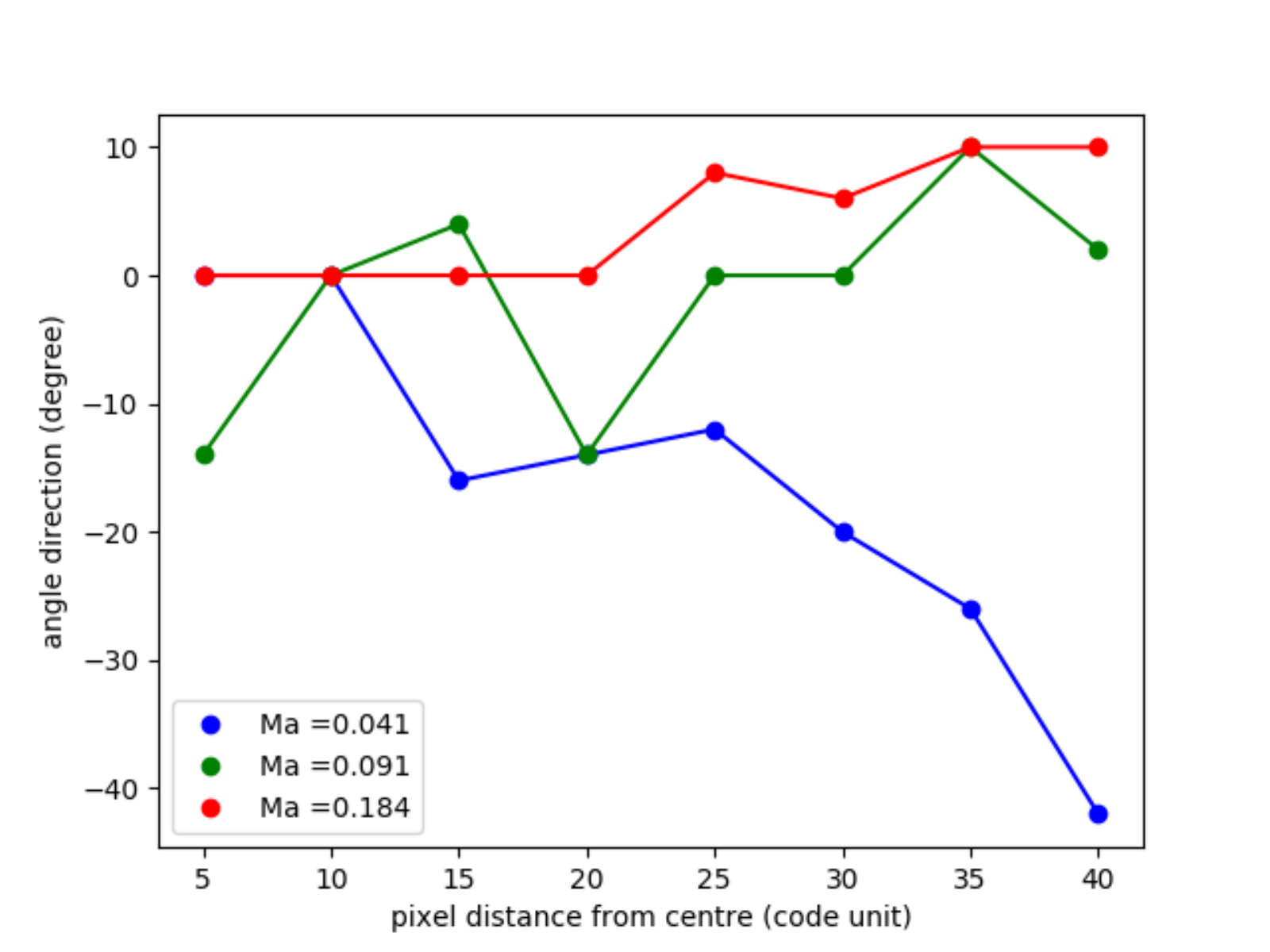}}
\caption{Showing two main properties of CFA to difference $M_a$ under resolution $480^3$ . Figure (a) shows the axis ratio to difference radius from the center. Figure (b) shows the variation degree difference to difference radius from the center. }
\end{figure*}

We first illustrate the effect of scale-dependent anisotropy in our numerical cube with lower resolution (\(480^3\)). Fig. \ref{fig:CFAdis0} shows how the shape and orientation of the correlation function anisotropy in Ms0.4Ma0.04 is changing with respect to length. One can directly see that while the numerical cube is somewhat anisotropic in all scales visually, both the axis ratio and the orientation are changing when one steps away from the center of the ellipses. On the left of Fig. \ref{fig:CFAdis1} shows a clearer effect with a scatter plot from three numerical simulations Ms0.4Ma0.04, Ms0.8Ma0.08 and Ms1.6Ma0.16, illustrating how the pixel distance (defined as the distance from the common center for the anisotropic ellipses) to the relative angle that the smallest anisotropy elongates to. 

\begin{figure*}[t]
\centering
\label{fig:CFAdis2}
\subfigure[]
{\includegraphics[width=0.49\textwidth]{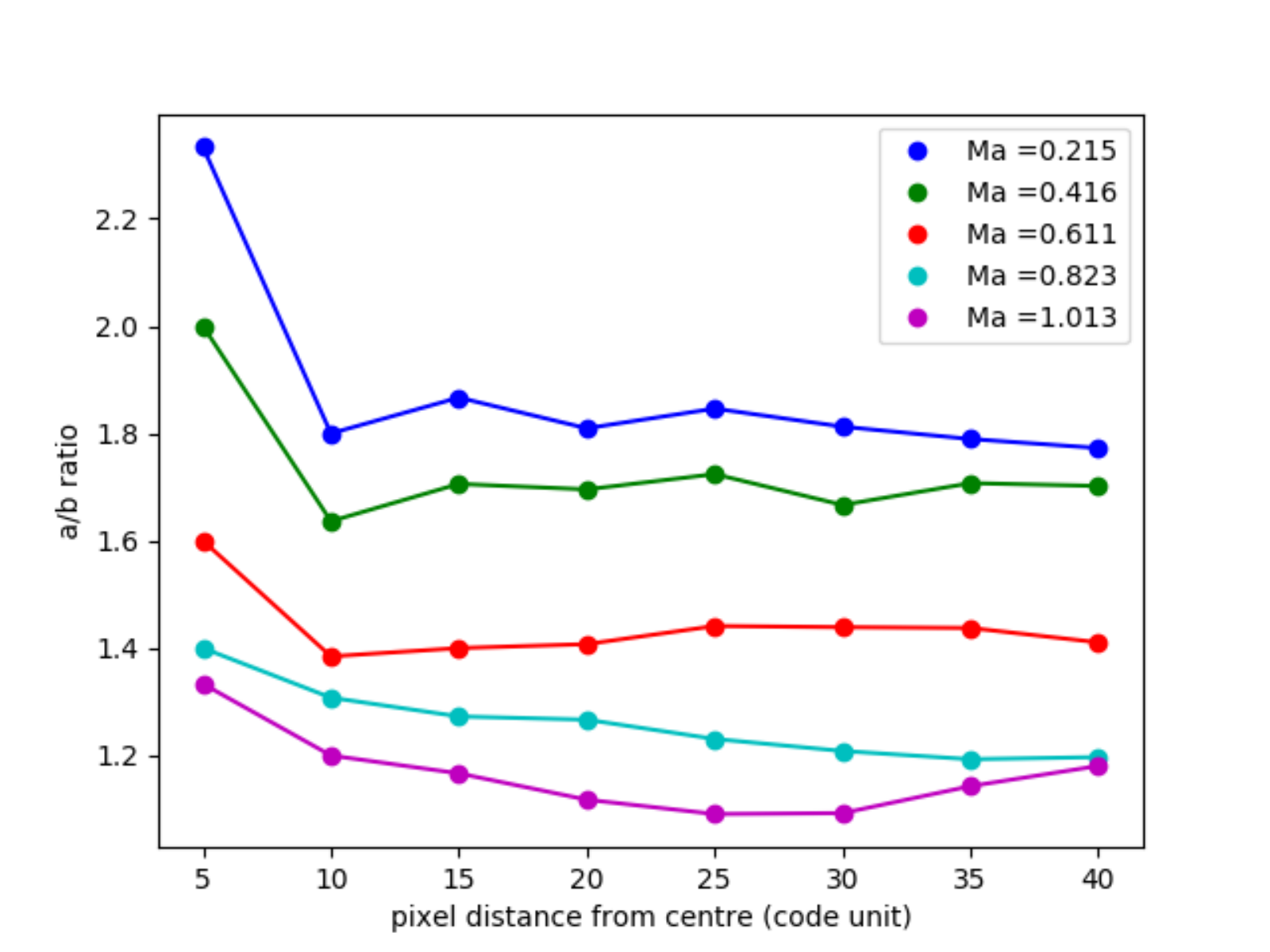}}
\subfigure[]
{\includegraphics[width=0.49\textwidth]{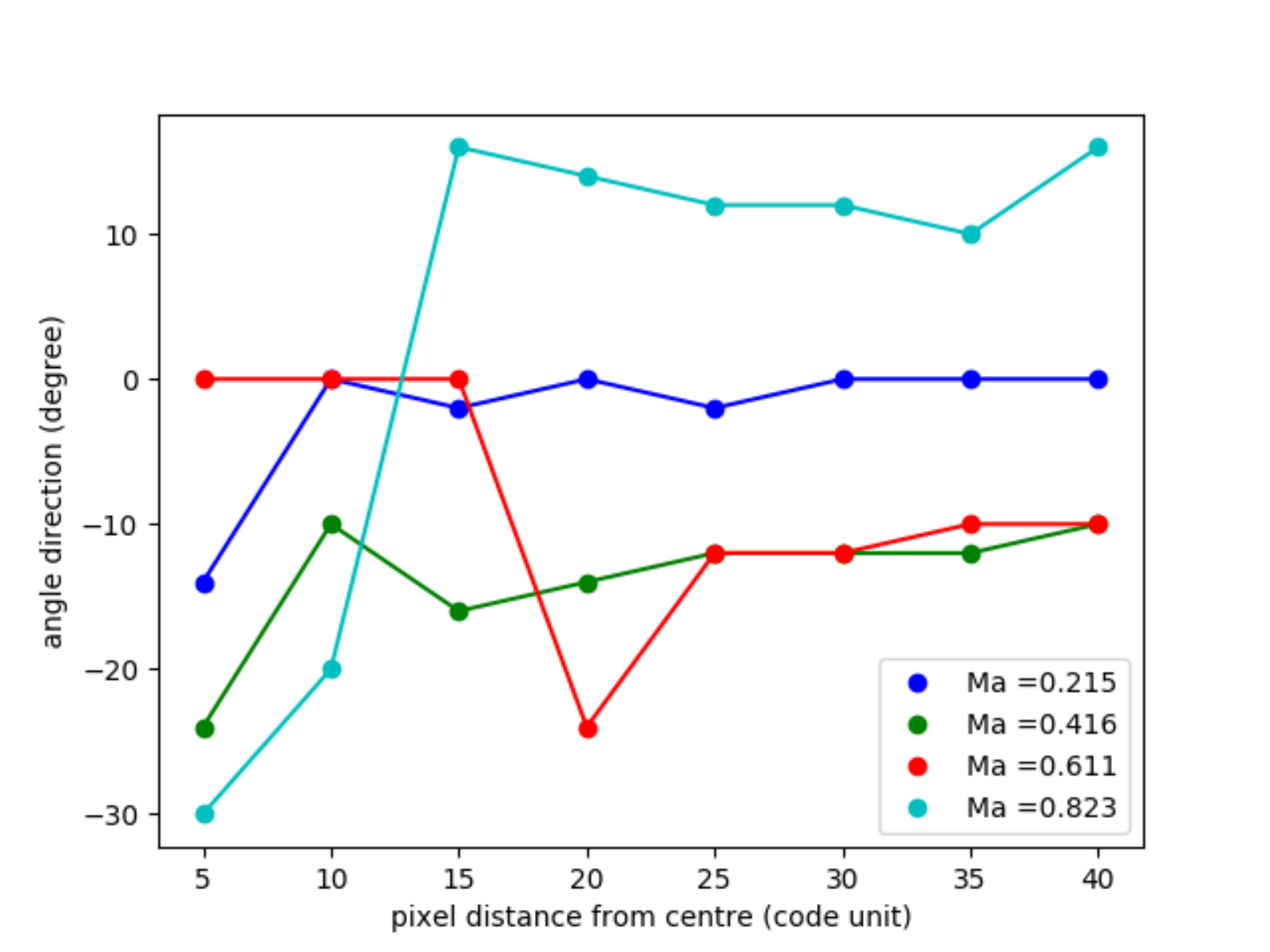}}
\caption{Showing two main properties of CFA to difference $M_a$ under resolution $792^3$ . Figure (a) shows the axis ratio to difference radius from the center. Figure b shows the variation degree difference to difference radius from the center.}
\end{figure*}

We can also try to quantify this effect with the major/minor axis ratio (a/b ratio). The right panel of Figure \ref{fig:CFAdis1} shows how the a/b ratio varies with respect to the pixel distance in our three selected simulations Ms0.4Ma0.04, Ms0.8Ma0.08 and Ms1.6Ma0.16. We can see an unexpected effect regarding the resolution: The smaller $M_A$ is actually being less anisotropic! What if we increase the resolution and simulate cubes with appropriate $M_A$ that its scale $L_{inj}M_A^2$ is within the inertial range. On Fig. \ref{fig:CFAdis2} shows how the cubes from the same code with higher resolution would behave. One can see from the trend for a/b axis ratio and orientation oscillations are closer to theoretical expectations \citep{EL05}, that a/b ratio is decreasing with respect to $M_A$, and orientation is more stable. This illustrates the resolution of the map is critical for the CFA study 

\subsubsection{Disappearance of elliptical anisotropy}

\begin{figure*}[t]
\centering
\label{fig:CFAdis3}
\includegraphics[width=0.98\textwidth]{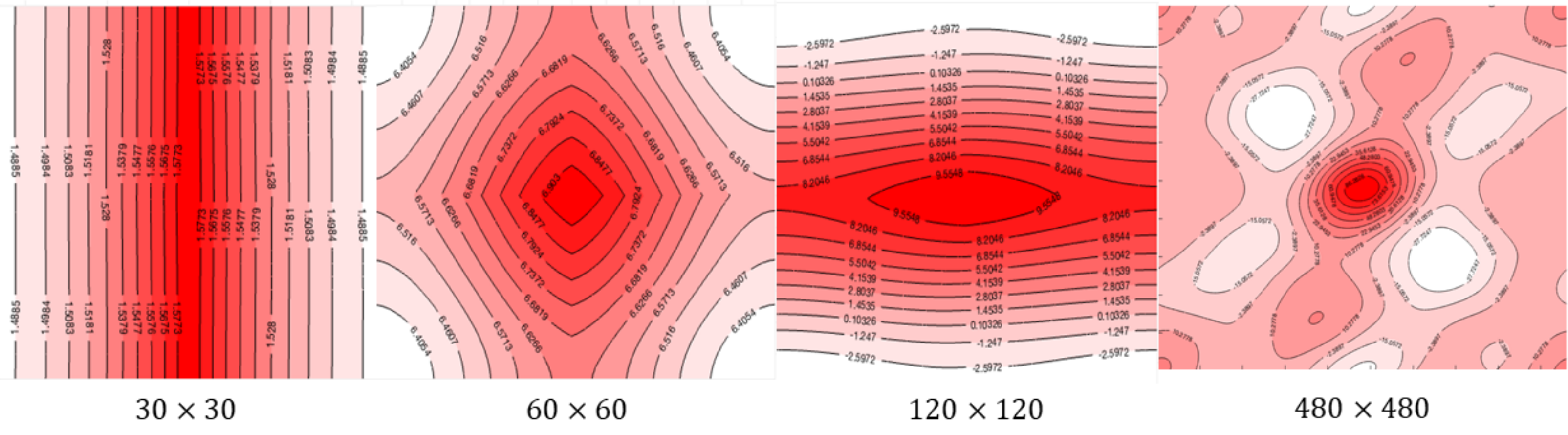}
\caption{The variation of correlation function anisotropy shapes with respect to block size for the cube Ms0.4Ma0.04. We draw contours to specify the isocontours. }
\end{figure*}

The resolution problem not only can distort the shape of the anisotropies in different scales, but also destroy the prominent elliptical shape, especially when performing the sub-block CFA analysis (Eq \ref{eq:obcf}). Figure \ref{fig:CFAdis3} shows how the shape of anisotropy is changed when one selects a different size of a partial region from the same cube Ms0.4Ma0.04. While the mean magnetic field stays pretty constant throughout the whole region, both the direction and the shape of contours change dramatically from elliptical to straight lines when one decreases the sampling size. 

It is worth noting that some isocontours in Fig \ref{fig:CFAdis3} show strip-like structures and, intuitively, the anisotropy is considered indeterminable in the region of interest. One possibility is that the measured region only contains small-scale turbulence in numerical box, which is contributed mostly from the dissipative ranges, so that the velocity motions in such a small region are similar to a noise-like environment. The effect of small area for CFA is similar to the uniform distribution for velocity gradient distribution we see in very small block size \citep{YL17a}, notifying that the samples (Both VGT and CFA) inside the region of interest is simply statistically not adequate for any kinds of anisotropy/magnetic field estimate. As the resolution decreases, the number of strip-like isocontours increases significantly, causing the result to be indeterminable for all scales of interest.

\subsection{Dependence of channel resolution in PCA}

While PCAA is a very powerful tool in extracting spectrum properties through a relatively simple statistical pipeline, there are concerns about its consistency and arbitrariness when applying to both synthetic and real observation. Two very important questions would be : 1. What is the minimal velocity channel number for PCAA and 2. What is the optimal number that one can pick for the ACF analysis? They are both crucial as how important a particular velocity spectral line eigenvector is depended on both the channel resolution and the number of biggest eigenvalues that are picked when fitting the $\alpha$ values.

\begin{figure*}[t]
\centering
\includegraphics[width=0.49\textwidth]{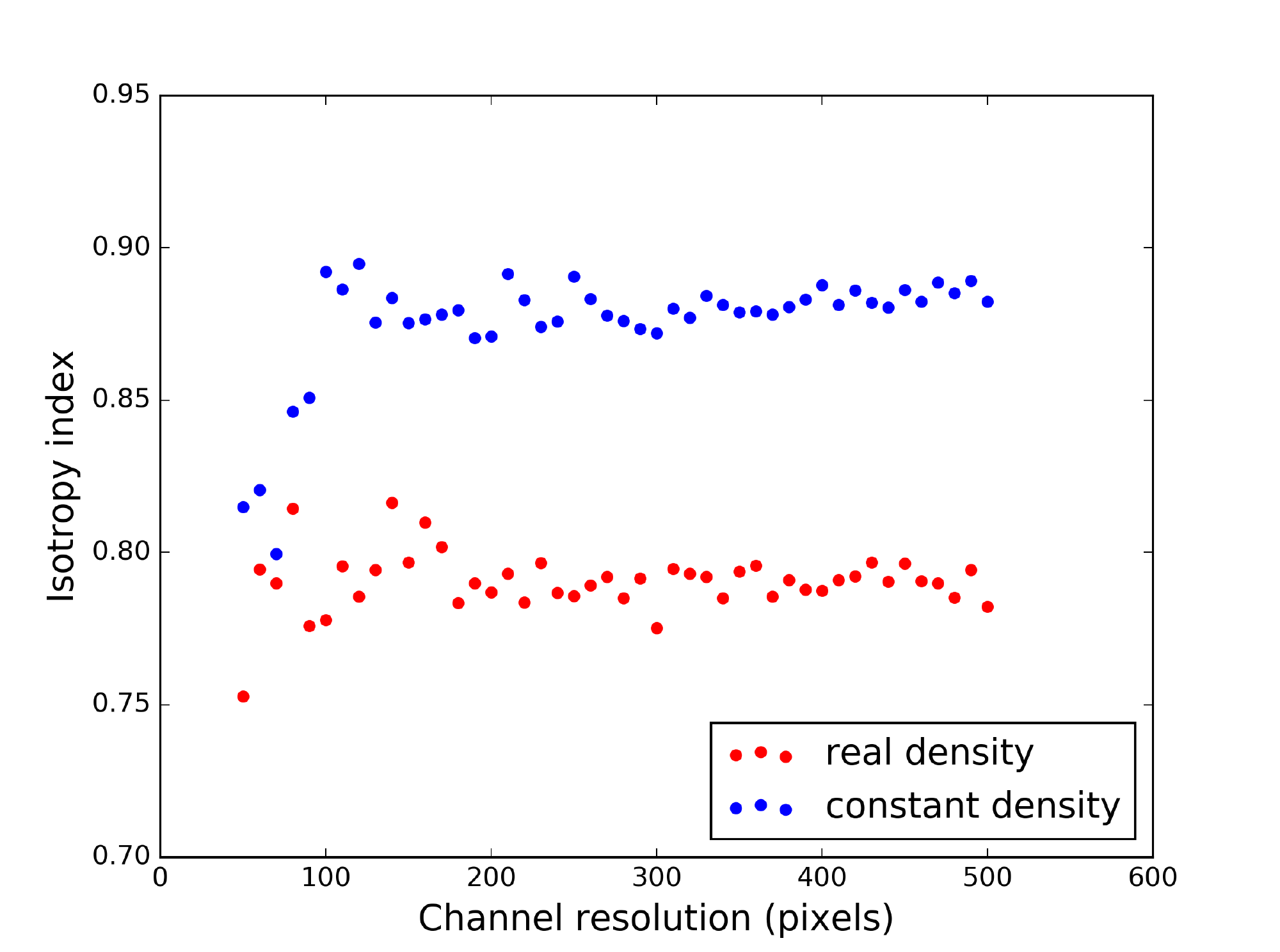}
\includegraphics[width=0.49\textwidth]{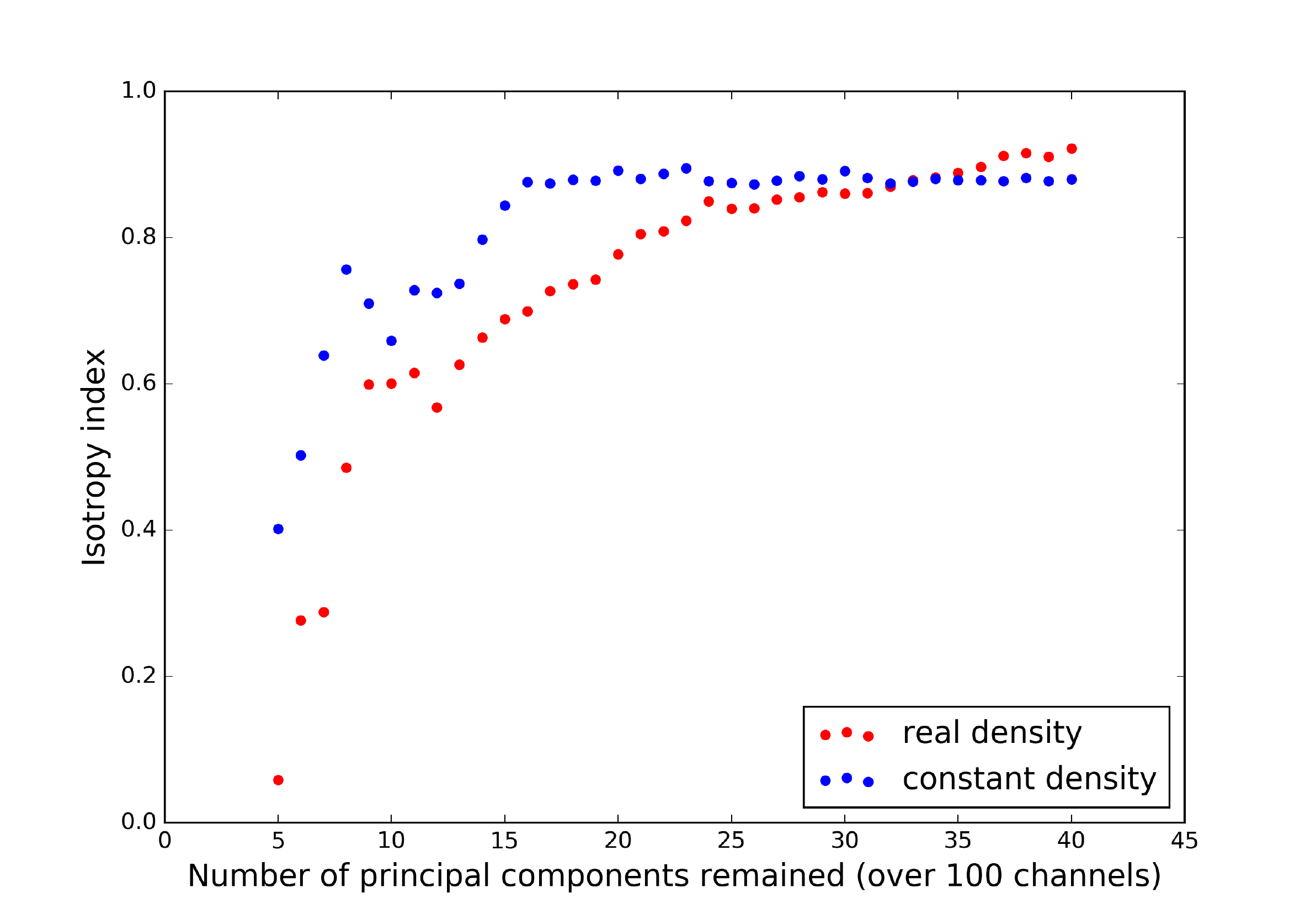}
\caption{\label{fig:PCAreflux}{\toreferee The plot shows how the isotropy index (y-axis) varies with respect to channel resolution (Left figure, x-axis) and the number of principal components remained (Right figure, x-axis)}.}
\end{figure*}

We first illustrate how the channel resolution would change the answer of PCAA. The left figure of Fig.\ref{fig:PCAreflux} shows a relation of isotropy index to the channel resolution for both constant and real density PPV cubes (See Eq (\ref{eq:nppv})) on the cube Ms1.6Ma0.528 by using 10 eigenvalues. One can see that the isotropy index fluctuates dramatically when the channel number is less than $200$ pixels and stays constant afterward. This indicates the method of PCAA has a more significant error if the velocity channel resolution is not enough. In particular, our test shows an approximately $~20\%$ error for constant density case and $~12\%$ for real density case in terms of the isotropy index. If one converts the isotropy index back to the angle, a larger error is expected. In this test, we did not include the noise produced by the instruments (which is very common in observational spectroscopic cubes). However, due to the nature of PCAA pipeline, the noise only contributes to the velocity spectra modes with small eigenvalues. 

We also test whether there is a way to search for the optimal number of principal components for PCAA analysis. In previous literature, only a handy of modes are used in PCAA (e.g. \citet{BH02} used 8 modes, and \cite{2016ApJ...818..118C} use 3 to 12). We use a $n_v=100$ cube to test how the change in the number of eigenvalues picked for PCAA would change the anisotropy measurement.  Fig. \ref{fig:PCAreflux} (right panel) shows the variation of isotropy index to the number of principal components we used for PCAA analysis, for both the actual turbulent density field (i.e.  denoted as "real") and a constant density field. One can see a significantly larger variation when the number of principal components is smaller than $20$. For instance, the isotropy index changes from $~0.05$ to $0.8$ when the number of principal components changes from $5$ to $15$ for the real density case. Relatively, the constant density PPV cube is more robust  when the number of principal components is changing, but the variation is still large  until component 20. 

One may argue whether the use of an intensity threshold for the higher eigen-projections may provide more stability to the isotropy index. However, the extra dependency on the intensity threshold will also increase the difficulty for observers to find the best combination of channel number, the number of components and the threshold value to analyze the result with PCAA.

\section{Summary}
\label{sec:conclusion}
In this paper, we studied three different techniques that have been suggested in the literature for tracing magnetic field. We improved these techniques and compared
them with each other. In particular, for the CFA technique we suggested and successfully tested a way of finding the direction of the magnetic field more accurately as well as a way of calculating correlation functions {\toreferee quickly}. We also suggested the sub-block averaging technique for the PCAA. {\toreferee Our finding} can be summarized as:
\begin{enumerate} 
\item Velocity Gradient Technique is {\toreferee a more superior technique compared to CFA and PCAA in tracing magnetic fields.} 
\item Correlation Function Anisotropy faces several issues when the block size is small. {\toreferee I}n particular, the anisotropies may be distorted, multi-centered or the contours are not closed. That significantly affects the determination of the direction of anisotropy, thus magnetic field. Poor resolution may also hinder the CFA technique from correctly determining the Alfv\'en Mach number.
\item Principle Component Analysis provides a method for extracting the most important velocity components in a PPV map. However, the detection of anisotropy strongly depends on the quality of spectroscopic cubes and the number of components that are being analyzed. We report only a weak dependence on $M_A$ and no dependence on $M_S$. With the block-averaging technique applied, we show that VGT has a significant advantage compared to PCA for finding magnetic field detections.
\end{enumerate}

\noindent{\bf Acknowledgments} The research activities of the Observational Astronomy Board at the Federal University of Rio Grande do Norte (UFRN) are supported by continuous grants from the Brazilian agencies CNPq and FAPERN and by the INCT-INEspaço. B.L.C.M. acknowledges a PDE/CAPES fellowship. A.L. acknowledges the support the NSF grant DMS 1622353 and AST 1715754, a Distinguished Visiting Professor  PVE/CAPES appointment at the Physics Graduate Program at UFRN, Natal, Brazil and a Visiting Professor Fellowship of the UFRN.  B.B. acknowledges support from the Institute for Theory and Computation (ITC) fellowship at the Harvard-Smithsonian Center for Astrophysics. We acknowledge Rachel and Vera for their kindly support of our numerical calculation. J.C.'s work is supported by the National Research Foundation of Korea Grants funded by the Korean Government (NRF-2016R1A5A1013277). {\toreferee We acknowledge Jeff Godsey from the Writing Center of UW-Madison for pointing language improvements on our paper.}

\software{Julia, Matlab, Python}


\appendix
\section{The FFT open-boundary cross-correlation method}
\label{sec:FFT}
Computationally the cross-correlation function is pretty expensive with computer complexity of $O(N^2)$. This disables scientists to study the statistical behaviour, often force the calculation of anisotropy to be truncated to small scales only. The reason behind is because the usual Fourier Transform method (Eq \ref{eq:fftc}) is not valid in the case when we select partial region for CFA analysis. In the following, based on the formulation from Hockney \& Eastwood we explain how one can compute the open boundary problem with similar treatment as in Eq. \ref{eq:cf}, which facilitates the method of CFA to observational maps ad also sub-block studies in parallel of VGT.

In the following treatment we shall interchange integrals $\int dx$ and summation signs $\sum_{x}$ freely to address the feature that the numerical data (both synthetic and observational) are discrete and having finite resolution.

Assuming we have a piece of complex numerical data $C(r)$ and we would like to obtain its correlation function:
\begin{align}
\label{eq:apcf1}
CF(r) = \int dr' C^*(r')C(r+r')
\end{align}
where the sign $*$ means complex conjugate. Using the definition of Fourier transform and assuming the functions are all $C_2$ converging, we have
\begin{align}
\label{eq:apcf2}
CF(r) 
&= \frac{1}{L^2}\int dr' dk_1 dk_2 C^*(k_1) e^{-ik_1(r'+r)} C(k_2) e^{ik_2r'}\\
&= \frac{1}{L}\int dk C^*(k)C(k) e^{-ikr}
\end{align}
where L is the normalization constant. This formula is essentially Eq \ref{eq:fftc}. 

Hockney propose a novel way of tackling the common convolution problem in computational physics especially to tackle Newtonian Gravity using the following implementation, which has been tested in our code in accelerating the open space discrete Poisson calculation from $O(N^2)$ to $O(NlogN)$. Here we follow their idea and calculate the cross-correlation counter-part following the Appendix of Ryne 2011 (arXiv:1111.4971). Suppose one is interested in the discrete cross-correlation
\begin{align}
f_i=\sum_{j=0}^{n-1} g_j^* h_{i+j}
\end{align}
where $i=0,1,2,...,m-1$. Define the zero-pad sequence of g which has the length of $N > m+n$
\begin{align}
\label{eq:Gdef}
G_j = \begin{cases}
g_j, & \text{if} \ j=0,1,2,...n-1\\
0, &\text{if} \ j=n+1,..,N
\end{cases}
\end{align}
Similarly, we can define the periodic cross-term
\begin{align}
H_k = \begin{cases}
h_k & \text{if}  j=0,1,2,...,m+n-1\\
0, & \text{if}  j=m+n,...,N\\
h_{mod(k,N)}, & \text{otherwise}
\end{cases}
\end{align}
The above is the basis the padding strategy as shown in Fig \ref{fig:HEps}.

The analogous summation formula for cross-correlation is similar to Eq. (A4) of Ryne (2011), assuming $W=e^{-2\pi{}i/N}$
\begin{align}
F_j &= \frac{1}{N} \sum_{k=0}^{N-1} W^{-jk} (\sum_{l=0}^{N-1} G_l^* W^{-lk})  (\sum_{l=0}^{N-1} H_l W^{lk}) \qquad\qquad 1 \leq j \leq N
\end{align}
And we already know Eq. \ref{eq:apcf2} the above formula will provide, in summation form and with the consideration of zero pads in Eq \ref{eq:Gdef}:
\begin{align}
F_i &= \sum_{i=0}^{N-1} G_i^* H_j = \sum_{i=0}^{m+n-1} g_i^* h_j
\end{align}

For the specific case that we are interested, $g=h=C$ and having the same size. The minimal number that satisfy the condition $N > m+n \sim 2n$ is $N=2n+1$. Noticing multidimensional FFT in rectangular case is orthogonal, we therefore arrive with the pictorial description as in Fig \ref{fig:HEps}.

{\toreferee One might question whether the appearance of bad pixels might alter the result we showed in the main text. We therefore perform a simple "hole-punching" test on our existing data. We select a centroid map C from the cube H0 (See Table \ref{tab:highres}) and randomly set a certain percentage of the pixels to be NaN. We then {\it directly} use the open-boundary FFT method on the centroid map with different block size to see whether the detected CFA orientation and axis ratio are changed after we zero the NaN pixels. Fig \ref{fig:punchhole-test} shows the change of axis ratio and orientation when we set a certain percentage of the data to be NaN and then perform CFA using the open-boundary FFT method. We see that even $40\%$ of the data is punch out we still have approximately the same predictions on the axis ratio or major-axis orientation, which suggested that the current open-boundary FFT method would still be robust to real data which will usually carry a number of empty pixels.

\begin{figure*}[t]
\centering
\label{fig:punchhole-test}
\subfigure[]
{\includegraphics[width=0.49\textwidth]{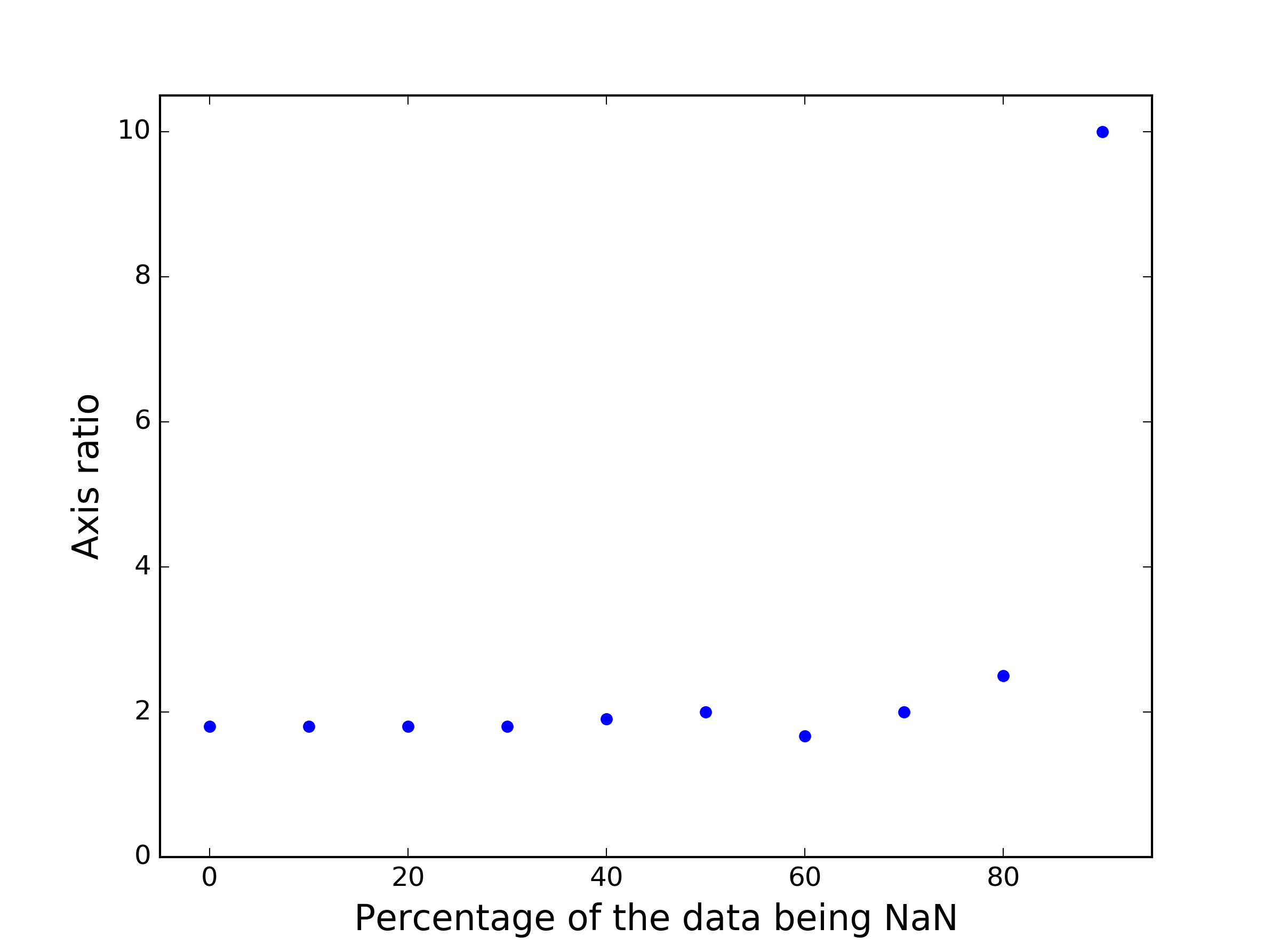}}
\subfigure[]
{\includegraphics[width=0.49\textwidth]{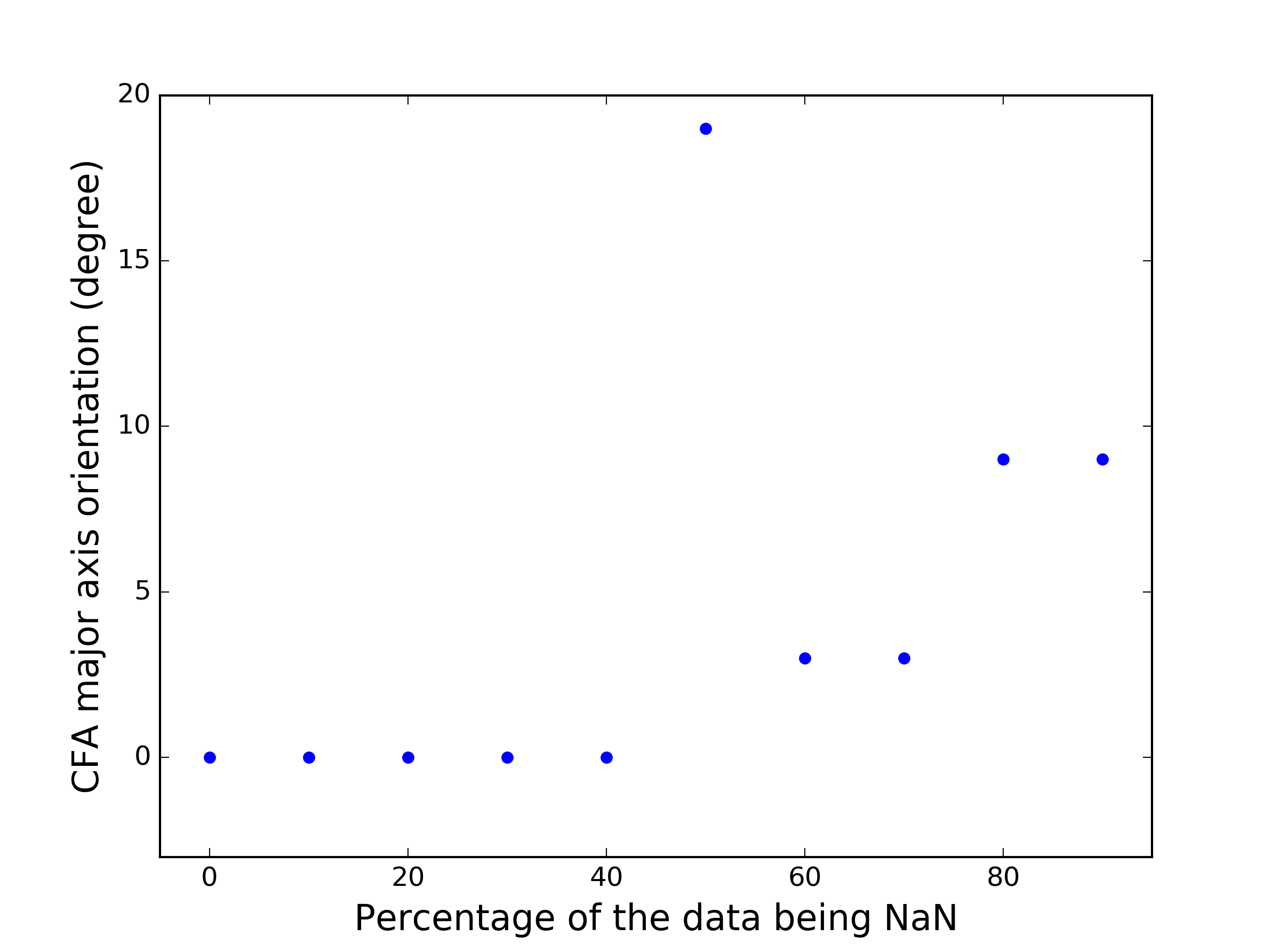}}
\caption{\toreferee Two panels showing the variance of axis ratio (a/b,left) and orientation of magnetic field as predicted by the CFA (right) when a certain percent of data is set to be NaN. }
\end{figure*}

}

\section{anisotropy}
A correct anisotropy can be revealed in a local frame of reference, which is defined with respect  to the local mean magnetic field. However, it is possible to observe a (fake) scale-dependent anisotropy in the global frame of reference, which is aligned with the mean magnetic field. For the sake of simplicity, we assume driving is isotropic throughout this Appendix.

If turbulence is sub-Alfvenic, it is easy to understand why we observe a (fake) scale-dependent anisotropy
in the global frame of reference. First, we note that large-scale structures are isotropic because driving is isotropic. Second, we note that structures measured in the global frame of reference are anisotropic on very small scales due to field-line wandering on large scales. The anisotropy in the global frame of reference is scale-independent on very small scales and  of order $l_{\|}/l_{\perp} \sim B_0/b_l \sim 1/M_A$, where $l_{\|}$ and $l_{\perp}$ are parallel and perpendicular size of eddies, respectively, and $M_A$ is the Alfven Mach number. Third, since large scales structures are isotropic and very small-scale ones are anisotropic, there should be transition scales, on which anisotropy is scale-dependent. Note that the scale-dependent anisotropy on the transition scales in the global frame of reference
is different from the true anisotropy that can be revealed in a local frame of reference.

\begin{figure*}[t]
\centering
\label{fig:app}
\includegraphics[width=0.99\textwidth]{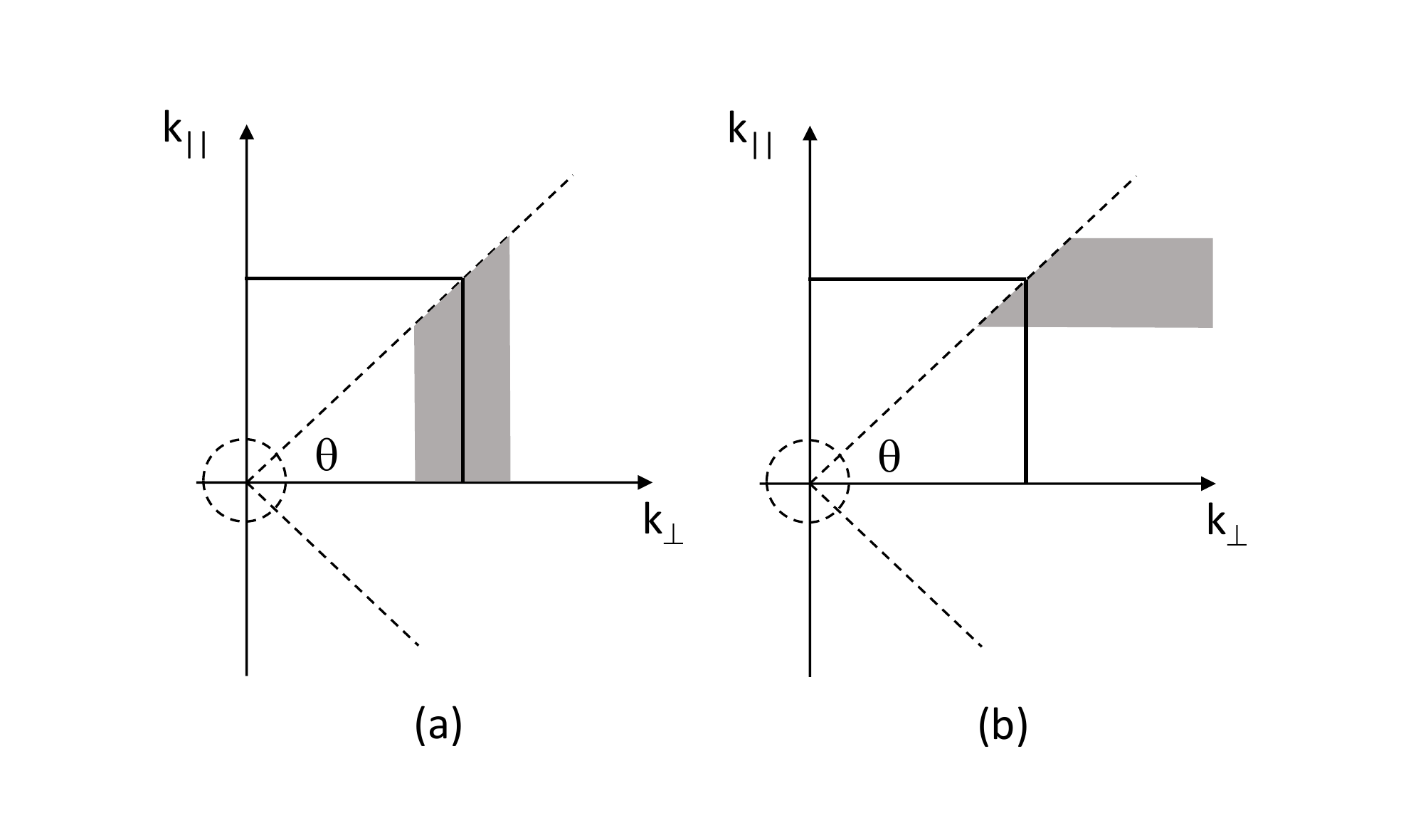}
\caption{Structure functions for the directions perpendicular (SF$_2(r_\perp, 0)$; left panel)  and parallel (SF$_2(0, r_{\|})$; right panel) to the mean magnetic field. Here $r_\perp \propto 1/k_\perp$ and $r_\| \propto 1/k_\|$.}
\end{figure*}

Even if $b_L/B_0 \sim M_A \sim 1$, we can have a (fake) scale-dependent anisotropy near the energy injection scale $L$ in the global frame of reference. Suppose that we try to reveal anisotropy using the second-order structure function
\begin{equation}
\textbf{SF}_2(r_\perp , r_\| ) = <|\textbf{A}( \textbf{x}+\textbf{r} )-\textbf{A}( \textbf{x})|^2>_{avg.~over  ~\textbf{x}},
\end{equation}

where \textbf{A} can be either velocity or magnetic field, $r_\perp$  and $r_\|$ are components of  the separation vector \textbf{r} perpendicular and parallel to the mean magnetic field, respectively. 
If contours of SF$_2(r_\perp , r_\| )$ are isotropic, we can say structures are isotropic (see Cho \& Vishniac 2000). If $b_L/B_0 \sim 1$, we expect that small-scale structures are isotropic and SF$_2(r_\perp, 0)$ = SF$_2(0,r_\|)$. Note, however, that roughly speaking the second-order structure function represent power near the scale of interest. For example, SF$_2(r_\perp, 0)$ represents power near the (perpendicular) scale $r_\perp$, which is approximately equal to the power in the shaded area in Figure \ref{fig:app}(a). Similarly, SF$_2(r_\perp, 0)$ is approximately equal to the power in the shaded area in Figure \ref{fig:app}(b). Although we will not  show it rigorously, Figure \ref{fig:app} clearly tells us that SF$_2(r_\perp, 0)$ $>$ SF$_2(0,r_\|)$, which means that structures look anisotropic. The fact that SF$_2(r_\perp, 0)$ $>$ SF$_2(0,r_\|)$ implies that contours of SF$_2(r_\perp , r_\| )$ are elongated along the direction parallel to the mean magnetic field. Note that this kind of anisotropy appears on sufficiently small scales. Now, the situation is similar to that of sub-Alfvenic turbulence: large scales are isotopic due to isotropic driving and small scales are anisotropic as we have shown above. Therefore, there should be transition scales, on which we observe a (fake) scale-dependent anisotropy.

\section{Term and Abbreviations}
\begin{center}
\begin{longtable}{|l|l|}
\caption{Term and Abbreviation used in the paper.} \label{tab:short} \\

\hline \multicolumn{1}{|c|}{\textbf{Abbreviation}} & \multicolumn{1}{c|}{\textbf{Term}} \\ \hline 
\endfirsthead

\multicolumn{2}{c}%
{{\bfseries \tablename\ \thetable{} -- continued from previous page}} \\
\hline \multicolumn{1}{|c|}{\textbf{First column}} & \multicolumn{1}{c|}{\textbf{Second column}}  \\ \hline 
\endhead

\hline \multicolumn{2}{|r|}{{Continued on next page}} \\ \hline
\endfoot

\hline \hline
\endlastfoot
AM  & Alignment Measure \\
ACF & Autocorrelation Function \\
CFA & Correlation Function Anisotropy    \\ 
VGT & Velocity Gradient Technique  \\
PCA & Principle Component Analysis  \\
PCAA & Principal Component Analysis of Anisotropies \\
PPV & Position-Position-Velocity  \\
pPPV & Partial Position-Position-Velocity  \\
pSIG & Polarised Synchrotron Intensity Gradient\\ 
VGT & Velocity Gradient Technique \\
VCG & Velocity Centroid Gradient \\
MHD & Magneto-hydro-dynamics \\
\end{longtable}
\end{center}

\end{document}